\documentclass[aps,prd,twocolumn, showpacs,showkeys,preprintnumbers,superscriptaddress,nobibnotes,floatfix,longbibliography,nofootinbib]{revtex4-1}
\pdfoutput=1
\usepackage{amsmath}
\usepackage{amsfonts}
\usepackage{amssymb}
\usepackage{mathrsfs}
\usepackage{graphicx}
\usepackage{color}
\usepackage{longtable}
\usepackage{multirow}
\usepackage{slashed}
\usepackage{epsf}

\usepackage[dvipsnames]{xcolor}

\usepackage{tikz}
\usetikzlibrary{shapes.geometric, calc}
\usepackage{colortbl}
\usepackage{placeins}
\usepackage{longtable}

\usepackage{color}
\definecolor{myred}{RGB}{255, 0, 0}
\definecolor{myblue}{RGB}{0, 0, 255}
\definecolor{mygreen}{RGB}{0, 128, 0}

\usepackage[colorlinks,
linkcolor=blue,
citecolor=blue,
urlcolor=blue
]{hyperref}

\begin{document}

\title{Light right-handed Smuons at the LHC:\\ Natural dark matter and $g_{\mu}-2$ in an unexplored realm of the pMSSM}

\author{Xiangwei Yin}
\email{yinxiangwei@cqu.edu.cn}
\affiliation{Department of Physics and Chongqing Key Laboratory for Strongly Coupled Physics, Chongqing University, Chongqing 401331, P.R. China}

\author{Tianjun Li}
\email{tli@itp.ac.cn}
\affiliation{CAS Key Laboratory of Theoretical Physics, Institute of Theoretical Physics,
	Chinese Academy of Sciences, Beijing 100190, China}
\affiliation{School of Physical Sciences, University of Chinese Academy of Sciences, No.~19A Yuquan Road, Beijing 100049, China}
\affiliation{School of Physics, Henan Normal University, Xinxiang 453007, P. R. China}

\author{James A. Maxin}
\email{hep-cosmology@pm.me}
\affiliation{Department of Chemistry and Physics, Louisiana State University, Shreveport, Louisiana 71115, USA}

\author{Dimitri V. Nanopoulos}
\email{dimitri@physics.tamu.edu}
\affiliation{George P. and Cynthia W. Mitchell Institute for Fundamental Physics and Astronomy,
	Texas A$\&$M University, College Station, Texas 77843, USA}
\affiliation{Academy of Athens, Division of Natural Sciences, 28 Panepistimiou Avenue, Athens 10679, Greece.}
\affiliation{Theoretical Physics Department, CERN, CH-1211 Geneva 23, Switzerland}



\begin{abstract}

We introduce an in-depth study of an unprobed pMSSM region offering a natural solution to dark matter. Since the 1980s this region has been referred to as the ``bulk'', consisting of sub 200~GeV neutralinos and right-handed smuons. The bulk satisfies recent muon $g_{\mu}-2$ measurements and sustains consistency with $all$ presently operating SUSY experiments and LHC constraints. Initial ingress into the bulk will arrive soon via the LUX-ZEPLIN 1000-day experiment. The ATLAS Collaboration at the LHC has confirmed that observation of these light right-handed smuon events can occur at the ongoing LHC Run 3 and forthcoming High-Luminosity LHC. Moreover, the future FCC-ee and CEPC circular colliders should handily observe the events.
\end{abstract}
\maketitle


\preprint{CERN-TH-2024-149}

\section{Introduction}{\label{Sec1}}

Supersymmetry (SUSY) provides a highly appealing framework for the exploration of new physics beyond the Standard Model (SM), and its investigation has been one of the primary objectives of the Large Hadron Collider (LHC). The minimal supersymmetric Standard Model (MSSM) with R-parity conservation represents the most economical extension of the SM, characterized by the minimal gauge group and particle content~\cite{Nilles:1983ge,Haber:1984rc}. See Ref.~\cite{MSSMWorkingGroup:1998fiq} for a review. However, this unconstrained MSSM (uMSSM) has over 100 free parameters, significantly reducing its predictive power.

In the quest for a more fundamental model at high energy scales, many embrace the grand unified theory (GUT) or string theory. In the MSSM, only soft SUSY-breaking terms that originate from a hidden sector interacting with the Standard Model through gravity are considered. Therefore, the soft SUSY-breaking terms are universal at high energies, for example, at the GUT scale, and satisfy certain boundary conditions.  These specific types of models are referred to as the constrained MSSM (cMSSM)~\cite{Kane:1993td}. In the full cMSSM, the soft SUSY-breaking mass parameters $M_0$, trilinear term $A_0$, and Gaugino mass $M_{1/2}$ are universal at the GUT scale. Moreover, the model includes \( \tan \beta \) and the undetermined sign of the SUSY breaking \( \mu \) term. These universal soft SUSY-breaking terms are related to low-energy parameters through renormalization-group equations (RGEs).

Some variant models relax the boundary conditions of the full cMSSM, such as dropping the assumption that Gaugino masses are universal at high-energy scales (NUGM)~\cite{Ellis:1984bm,Wang:2021bcx,Wang:2018vrr,Aboubrahim:2021xfi}. Thus, the bino mass \( M_1 \), wino mass \( M_2 \), and gluino mass \( M_3 \) can be different. Alternate models relax the assumptions of universality for the soft SUSY-breaking contributions to the Higgs masses, which differ from the universal scalar mass $M_0$, for instance, NUHM1~\cite{Baer:2004fu,Baer:2005bu,Ellis:2008eu,Ellis:2012nv,Buchmueller:2013rsa}, NUHM2~\cite{Ellis:2002wv,Ellis:2002iu,Buchmueller:2014yva}, and NUHM3~\cite{Baer:2021aax,Ellis:2024ijt}, etc.

In particle physics, there has been a long-standing discrepancy between the theoretical prediction and the experimental measurement of the muon's anomalous magnetic moment $a_{\mu}\equiv(g-2)_{\mu}/2$. The state-of-the-art experimental world average from BNL and Fermilab (BNL+FNAL) is~\cite{Muong-2:2021ojo,Muong-2:2023cdq} 

\begin{equation}
a_\mu^{\text{BNL+FNAL}}=116592059(22) \times 10^{-11}~,
\end{equation}
while the data-driven theoretical calculation yields

\begin{equation}
a_\mu^{\text{SM}}=116591810(43) \times 10^{-11}~.
\end{equation}
Combining the data-driven theoretical prediction and the experimental measurement, the deviation is 
\begin{equation}
\Delta a_\mu^{\text{BNL+FNAL}}=(24.9 \pm 4.8) \times 10^{-10}~.
\end{equation}
This well-known data-driven value notwithstanding, there exist some recent lattice calculations (BMW) yielding a smaller disparity of $\Delta a_{\mu}=(10.7\pm6.9)\times10^{-10}$~\cite{Borsanyi:2020mff,Kuberski:2023qgx}. 

Achieving a significant contribution to the muon $g_{\mu}-2$ anomaly in supersymmetric models clearly necessitates relatively light neutralinos and smuon masses, typically around a few hundred GeV. However, this is challenging to realize within the framework of the full cMSSM due to LHC experimental constraints that impose lower limits on the masses of gluinos and squarks above 2 TeV. These constraints render obstacles for universal boundary conditions to yield light electroweakinos and sleptons. Consequently, as previously mentioned, the universal boundary conditions can be relaxed on the Gaugino mass or scalar mass. This permits such models to be consistent with the experimentally observed value of the muon $g_{\mu}-2$, while also satisfying the mass constraints on supersymmetric particles.

Instead of the UV completion, we study here the more general phenomenological MSSM (pMSSM), which does not assume universal boundary conditions at the GUT scale and provides an excellent framework for studying low-energy SUSY phenomenology. In this paper, we investigate in the pMSSM a region that is referred to as the ``bulk'', recently systematically studied in Ref.~\cite{Yin:2023jhs}, and home of light neutralinos and light right-handed sleptons. Numerous prior analyses have studied light mass regions of the pMSSM parameter space in the context of light dark matter (DM)  and collider searches~\cite{Hooper:2002nq}, though to our knowledge this is the first analysis of the bulk with light right-handed smuons within the pMSSM. Recall that the bulk was the original ambition in the 1980s for SUSY discovery~\cite{Ellis:1983ew} due to its naturally light neutralinos and right-handed sleptons. However, something happened on the road to SUSY camelot. Each successive run of LEP, Tevatron, and LHC failed to unveil any tantalizing signals of SUSY, thereby pushing sparticle mass limits higher and higher. Indeed, the LHC has elevated gluino and squark constraints above 2~TeV, making it extremely challenging to theoretically model light neutralinos and right-handed sleptons given the heavy remaining part of the spectrum. This yielded heavier neutralinos and sleptons from simplified models in the ensuing years, which encouraged finely tuned methods to satisfy dark matter relic density observations, namely, the ``Higgs funnel'', ``sfermion coannihilation'', and the ``mixing scenario'' or ``well-tempered scenario''. So in essence, by formulating the bulk and studying its observable attributes, we are returning back to the dawn of the SUSY era and revitalizing the most natural and theoretically motivated domain for discovery.

Given the predicament of deriving light neutralinos and smuons in the cMSSM with present elevated LHC constraints, the pMSSM eliminates all universality, allowing the pMSSM to generate light neutralinos and smuons. An important point to emphasize here though is that our derivation of light neutralinos and right-handed sleptons in the pMSSM is no fluke occurrence. Both left-handed and right-handed sleptons occupy $SU(5)$ supermultiplets alongside squark masses, thus, RGE running binds heavy squark masses to heavy neutralinos and sleptons in constrained versions of the SUSY model space, such as the cMSSM, where $SU(3)$, $SU(2)$, and $U(1)$ Gaugino masses are universal. At a minimum, light neutralinos and right-handed sleptons require universality to be broken so that the $U(1)$ factor and right-handed scalar mass $m_{\Tilde{e}_{R}} = m_{\Tilde{\mu}_{R}}$ and $m_{\Tilde{\tau}_{1}}$ RGEs can run independently. The unconstrained pMSSM permits this autonomy, but it does not guarantee strong theoretical underpinnings as to why the light neutralinos and sleptons manifest. The study presented in this work will showcase derivation of the bulk in the pMSSM in the quest for naturally generated DM via light neutralinos, and hence light right-handed smuons. As a consequence, and indeed bonus, the bulk also naturally yields sufficient $a_{\mu}$ to produce the tentatively observed excess beyond the expected SM value. This is contrary to just some random scan of the pMSSM parameter space that haphazardly stumbles upon light neutralinos and sleptons but possesses no theoretical logic as to why.

Before proceeding here with our pMSSM analysis, we pause to note that there is a GUT model where light right-handed smuons can coexist with universality, namely, ${\cal F}$-$SU(5)$~\cite{fsu5old,fsu5}. In flipped $SU(5)$ with gauge symmetry $SU(5) \times U(1)_X$, we have a universal Gaugino mass for wino and gluino, but not for the bino due to the additional $U(1)_X$ gauge symmetry. Therefore, the bino can be very light and be the lightest supersymmetric particle (LSP). Moreover, the right-handed sleptons are singlet under $SU(5)$ and only charged under $U(1)_X$. Hence, the right-handed sleptons can be light naturally, and the right-handed smuon can be the next to the lightest supersymmetric particle (NLSP) if the masses for the right-handed sleptons are not universal. Consequently, we might have the following supersymmetry breaking scenario: the LSP neutralino with bino dominant component, the NLSP right-handed smuon, and the other supersymmetric particles which are relatively heavy. This scenario might explain the muon anomalous magnetic moment as well.

To uncover the bulk, the parameter space of the compressed mass region in the pMSSM was scanned, taking into account experimental constraints on sleptons, gluinos, squarks, rare B-meson decays, light Higgs boson mass, DM relic density, and both spin-dependent and spin-independent DM-proton scattering cross-sections. By selecting pure bino DM, avoiding resonance annihilation, and suppressing coannihilation, we uncover the bulk. This portion of the parameter space has not yet been excluded by collider experiments but holds promise for upcoming verification at the very early 1000-day LUX-ZEPLIN~\cite{LZ:2018qzl,LZ:2022lsv}, the current LHC Run 3 (LHC3), High-Luminosity LHC (HL-LHC), Future Circular Collider (FCC-ee)~\cite{FCC:2018byv,FCC:2018evy} at CERN, and Circular Electron Positron Collider (CEPC)~\cite{CEPCStudyGroup:2018ghi}. Notably, the bulk region naturally features very light neutralinos and right-handed smuons, which are consistent with the combination of BNL+FNAL measurements while also being in agreement with the CMD-3 discrepancy of $\Delta a_{\mu}=(4.9\pm5.5)\times10^{-10}$~\cite{CMD-3:2023rfe} that relies upon the $e^+ e^- \rightarrow \pi^+ \pi^-$ channel, which motivated recalculations of previous data, giving $\Delta a_{\mu}=(12.3\pm 4.9)\times10^{-10}$~\cite{Davier:2023fpl}. Hence, this region serves as an excellent parameter space for studying the muon $g_{\mu}-2$, and in the most favorable scenario, direct production of light right-handed smuons could be observed at the LHC3 and HL-LHC.

\begin{figure}[t]
    \centering
    \includegraphics[width=1.0\linewidth]{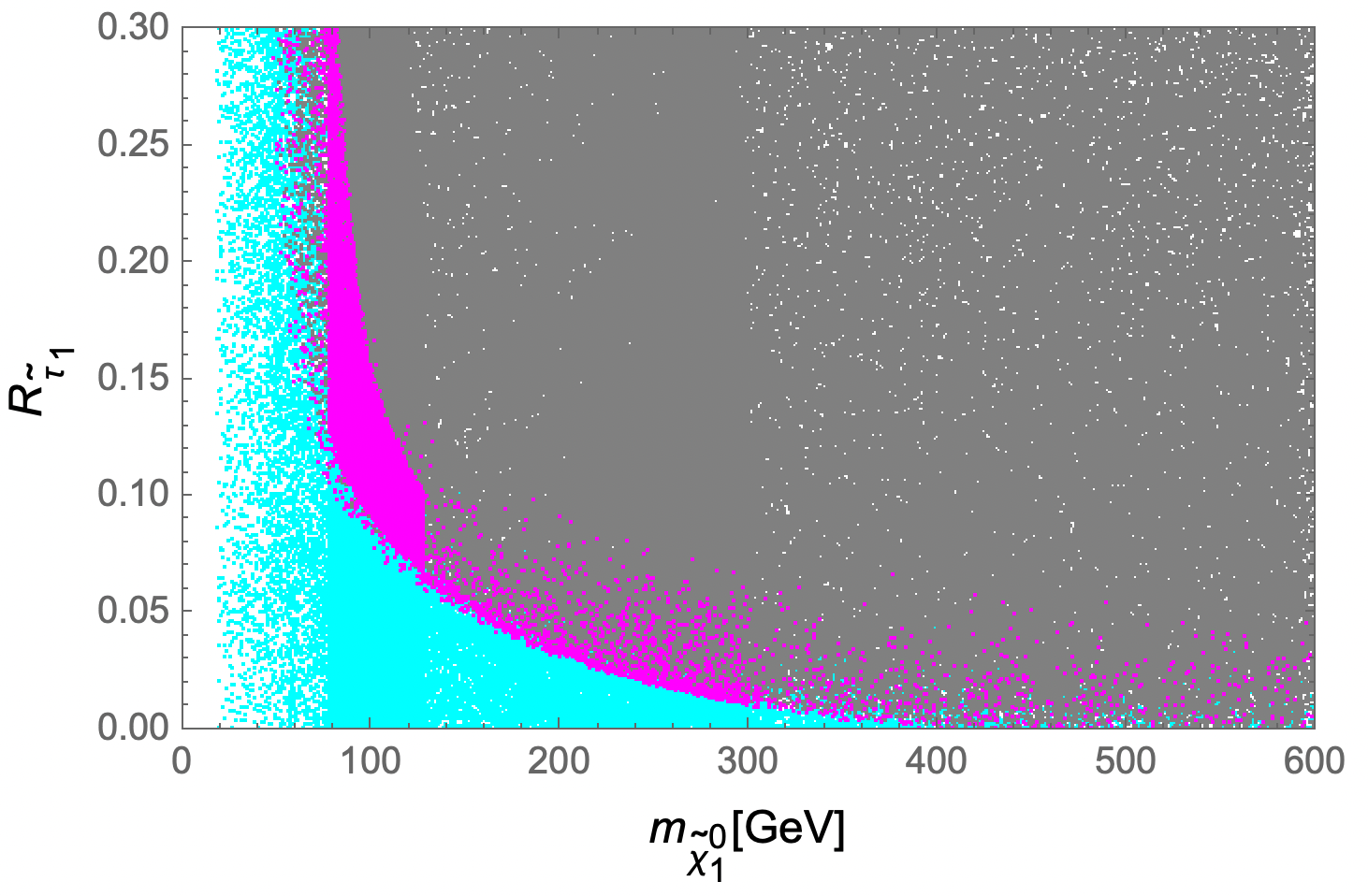}
    \caption{Bulk region in the pMSSM with $\mathcal{R}_{\Tilde{e}_{R}}> \mathcal{R}_{\Tilde{\tau}_{1}}$. Cyan, magenta, gray points correspond to undersaturated, saturated, oversaturated DM relic density.}
    \label{C4_F01_bulk}
\end{figure}

This paper is organized as follows: In Sec.~\ref{Sec2}, an introduction to the pMSSM model and bulk region is provided, highlighting that the bulk features a light neutralino and smuon, which can naturally enhance the muon $g_{\mu}-2$. Section~\ref{Sec3} contains a discussion of the experimental constraints and sequence of steps to uncover the bulk, and then Sec.~\ref{Sec4} presents the phenomenological results, with a concluding summary in Sec.~\ref{Sec5}.

\section{\MakeLowercase{p}MSSM}{\label{Sec2}}

The pMSSM is a more general model that consists of 22 free parameters. We replace $m_{H_{u}}^{2}$ and $m_{H_{d}}^{2}$ with $M_{A}$ and $\mu$ as inputs and choose the sign of the SUSY breaking parameter $\mu$ to be positive, sgn$(\mu) > 0$. The 22 free parameters are scanned within the following ranges:

\begin{eqnarray*}
   &20 ~\text{GeV}\leq M_{1}\leq 1000 ~\text{GeV} \\
   &1000 ~\text{GeV}\leq M_{2}\leq 5000 ~\text{GeV} \\
   &1200 ~\text{GeV}\leq M_{3}\leq 5000 ~\text{GeV} \\
   &M_{1} \leq m_{\Tilde{e}_{R}} = m_{\Tilde{\mu}_{R}}, m_{\Tilde{\tau}_{1}} \leq 3 M_{1} \\
   &2500 ~\text{GeV}\leq m_{\Tilde{q}},m_{\Tilde{Q}},m_{\Tilde{u}_{R}},m_{\Tilde{t}_{R}},m_{\Tilde{d}_{R}},m_{\Tilde{b}_{R}}    \leq 5000 ~\text{GeV} \\
   &700 ~\text{GeV}\leq m_{\Tilde{l}},m_{\Tilde{L}}\leq 2000 ~\text{GeV} \\
   &-5000 ~\text{GeV}\leq A_{u}, A_{d}, A_{e}, A_{t}, A_{b}, A_{\tau} \leq 5000 ~\text{GeV} \\
   &2 \leq {\rm tan}\beta\leq 65 \\   
   &1000 ~\text{GeV}\leq M_{A}, \mu \leq 6000 ~\text{GeV} \\
\end{eqnarray*}
where $m_{\Tilde{q}}$, $m_{\Tilde{u}_{R}}$, $m_{\Tilde{d}_{R}}$, $m_{\Tilde{l}}$, and $m_{\Tilde{e}_{R}} = m_{\Tilde{\mu}_{R}}$ are the first/second generation sfermion mass parameters and $m_{\Tilde{Q}}$, $m_{\Tilde{t}_{R}}$, $m_{\Tilde{b}_{R}}$, $m_{\Tilde{L}}$, and $m_{\Tilde{\tau}_{1}}$ are third generation sfermions. 

In the pMSSM, the lightest neutralino $\tilde{\chi}_1^0$ can serve as a DM candidate, and its eigenstate can be expressed as a linear combination of the bino ($\tilde{B}$), wino ($\tilde{W}$), and two neutral Higgsinos ($\tilde{H}^0_1, \tilde{H}^0_2$)~\cite{Ellis:1983ew}
\begin{equation}
\tilde{\chi}_1^0=N_{11} \tilde{B}+N_{12} \tilde{W}^3+N_{13} \tilde{H}_1^0+N_{14} \tilde{H}_2^0~.
\label{DMeq}
\end{equation}
Our objective is to identify regions within the pMSSM with relatively small neutralino and smuon masses that are consistent with the muon $g_{\mu}-2$ experimental results, while also satisfying constraints from collider experiments, DM detection, and other experimental observations. Free of the severe restrictions imposed by universal boundary conditions, the pMSSM can naturally accommodate lighter sleptons; however, experiments related to DM direct detection and relic density still impose stringent constraints. For instance, a bino LSP is tightly constrained by relic density requirements, while wino and Higgsino are limited by direct detection experiments.

The relic density of a binolike neutralino $\Tilde{\chi}^{0}_{1}$ typically tends to be excessively high, either due to the small couplings of the LSP, the heavy masses of the sparticles, or both \cite{Martin:1997ns}. The correct relic density can be achieved through four canonical approaches. Alternative methods, such as introducing gravitino or axino LSPs, are not discussed here.
The original approach is the bulk region~\cite{Ellis:1983ew}, where $\Tilde{e}_{R}$, $\Tilde{\mu}_{R}$, and $\Tilde{\tau}_{1}$ are light. In this scenario, the t-channel exchange of these sfermions is sufficient to reduce the relic density to the observed value. The second method is the Higgs funnel, where the binolike LSPs satisfy the mass relation $2m_{\Tilde{\chi}^{0}_{1}} \approx m_{A^{0}}$, $m_{h}$, or $m_{H^{0}}$. The third mechanism is the sfermion coannihilation region, where masses of the sfermions and LSP are nearly degenerate, within a few GeV. The last solution is the mixing scenario or well-tempered scenario~\cite{Arkani-Hamed:2006wnf}, where the LSP neutralino has enough wino or Higgsino component to significantly increase the annihilation cross section.

\begin{figure*}[htb]
  \centering
  \begin{minipage}{0.46\textwidth}
    \centering
    \includegraphics[width=\textwidth]{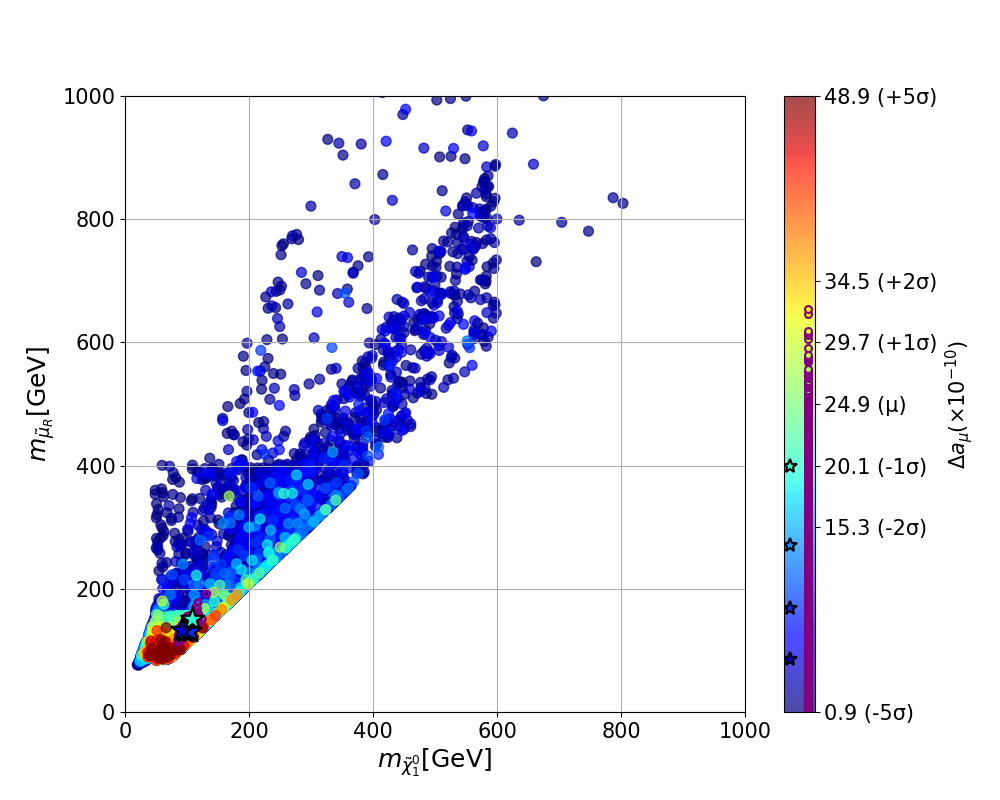}
    \par\smallskip
    \textbf{(a)}
  \end{minipage}
  \hspace{0.05\textwidth}
  \begin{minipage}{0.46\textwidth}
    \centering
    \includegraphics[width=\textwidth]{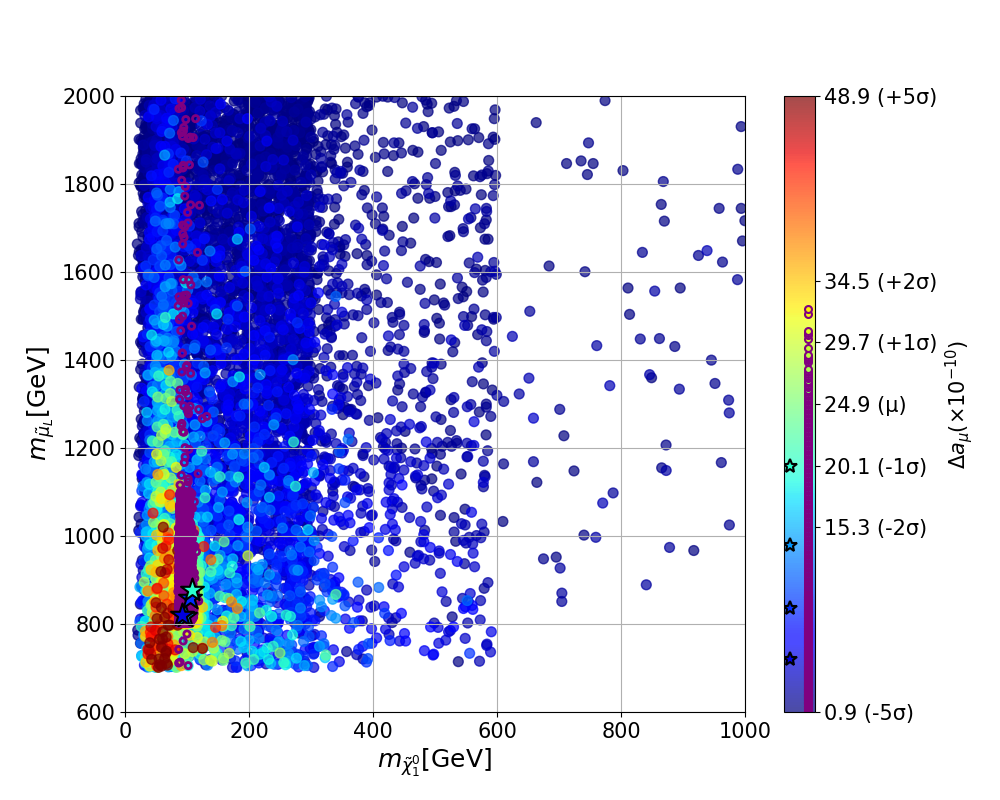}
    \par\smallskip
    \textbf{(b)}
  \end{minipage}
  \caption{Depiction of $m_{\Tilde{\mu}_{R}}$ {\bf (a)} and $m_{\Tilde{\mu}_{L}}$ {\bf (b)} versus $m_{\chi^0_1}$ in the pMSSM. Points in {\bf (b)} adhere to $m_{\Tilde{\mu}_{L}} > 700$ GeV for consistency with slepton LHC collider searches. Every solid point in both plot spaces are color coded to the $\Delta a_{\mu}$ color scale on the right side of each plot, with the exception of the open purple circles (heavily overlapped in the denser areas) identifying the bulk, both in the plot space and color scale. The stars with black outline denote four benchmark points for the bulk region given in Table~\ref{C4_T1_Benchmark}, and the interior color of each star correlates to the precise color assigned to that particular $\Delta a_{\mu}$ value per the color scale. All points satisfy the experimental constraints listed in Sec.~\ref{Sec3}, and the bulk points further comply with the bulk requirements itemized in Sec.~\ref{Sec3}.}
  \label{C4_F03_amuSmuonR}
\end{figure*}

The bulk region in No-scale $\mathcal{F}$-$SU(5)$ and the pMSSM was analyzed in Ref.~\cite{Yin:2023jhs}, and we expand upon the detailed parameter space in the pMSSM in this work. The anomalous magnetic moment of the muon arises from supersymmetric contributions, primarily at leading one-loop order from diagrams involving charginos and sneutrinos, as well as neutralinos and smuons~\cite{Lopez:1993vi,Moroi:1995yh,Martin:2001st}. Significant contributions to the muon $g_{\mu}-2$ primarily arise from regions where masses of supersymmetric particles are relatively light, around a few hundred GeV. Coannihilation and resonance annihilation require fine-tuning of the bino mass, whereas the well-tempered scenario faces more stringent constraints from direct detection experiments. Therefore, we regard the bulk region to be the most natural parameter space. In this region, it is feasible to have naturally light neutralinos and smuons, which can significantly enhance the contribution to the muon $g_{\mu}-2$. More importantly, in this compressed mass region, the right-handed smuon can be very light, around a few hundred GeV, just slightly beyond prior detection capabilities of LHC experiments. Via discussions with the ATLAS Collaboration as to whether the smuon search limits of Ref.~\cite{ATLAS:2019lff} can be extended to fully cover the light right-handed smuon parameter space, ATLAS did confirm that the light right-handed smuon can be comprehensively probed in the coming years at the LHC3 and HL-LHC.

\section{Analytical Procedure}{\label{Sec3}}
LHC constraints on the compressed mass parameter space have evolved to rather stringent levels, but these strong limits can be relaxed in the event the sparticle masses are nearly degenerate. Without considering R-parity violation or introducing new particles as DM candidates, the lightest neutralino remains absolutely stable and can serve as a DM candidate. We implement experimental constraints on the gluino mass, squark mass, B-meson decay $B_s^0 \rightarrow \mu^{+} \mu^{-}$, radiative flavor-changing b-quark decay $b \rightarrow s \gamma$, DM relic density, spin-dependent and spin-independent scattering cross-sections, and collider experiment limits on sleptons. The rigid constraints are as follows:

\begin{figure}[htbp]
    \includegraphics[width=1.0\linewidth]{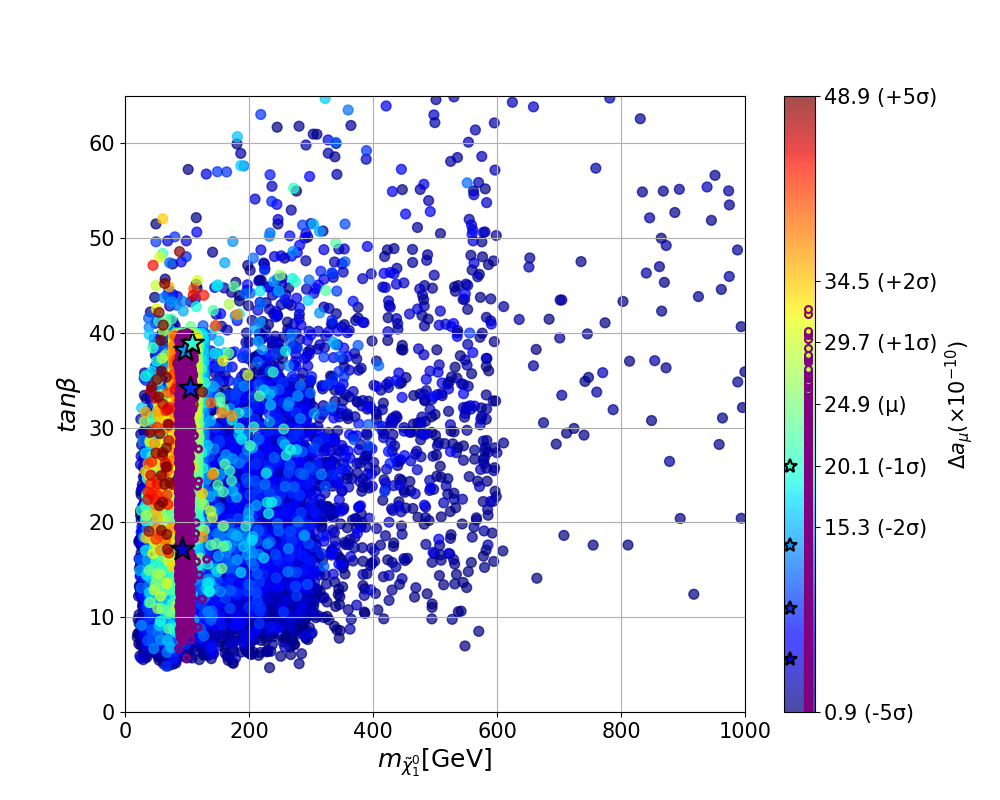}
  \caption{The $\tan \beta$ versus $m_{\chi^0_1}$ plane in the pMSSM. Notice that large $\tan\beta$ does enhance $\Delta a_{\mu}$ in the bulk. The same color scheme and constraints of Fig.~\ref{C4_F03_amuSmuonR} apply in this figure as well.}
  \label{C4_F05_amuTanb}
\end{figure}

\begin{itemize}
    \item Collider limits on masses of gluinos and the first two generations of squarks of $m_{\Tilde{g}}\gtrsim 2.2 ~\text{TeV} $,~$m_{\Tilde{q}}\gtrsim 2.0 ~\text{TeV}$~\cite{ATLAS:2017mjy,Vami:2019slp,CMS:2017okm}.
    \item Rare B-meson decay branching ratio of $1.6 \times 10^{-9}$ $\leq$ BR($B_{s}^0 \rightarrow \mu^{+}\mu^{-}$) $\leq$ $4.2 \times 10^{-9}$~\cite{CMS:2014xfa} and radiative flavor-changing b-quark decay branching ratio of $2.99 \times 10^{-4} \leq$ BR($b\rightarrow s \gamma$) $\leq 3.87 \times 10^{-4}$~\cite{HFLAV:2014fzu}.
    \item Attention to both experimental and theoretical uncertainties of the light Higgs boson mass, applied as 122 GeV$\leq m_{h} \leq$ 128 GeV ~\cite{ATLAS:2012yve,CMS:2012qbp,Slavich:2020zjv,Allanach:2004rh}.
    \item Spin-independent DM-proton scattering cross-section from the LUX-ZEPLIN (LZ)~\cite{LZ:2018qzl,LZ:2022lsv} experiment, detection potential of the future LZ 1000-day experiment~\cite{LZ:2018qzl}, and spin-dependent scattering cross-sections from PICO-60 $C_{3}F_{8}$~\cite{PICO:2019vsc} and IceCube~\cite{IceCube:2016dgk}.
    \item DM relic density constraints within the Planck 2018 experiment 5$\sigma$ range of $0.114 \leq \Omega_{\rm DM}h^{2} \leq 0.126$~\cite{Planck:2018nkj}. Points within this range we shall refer to as saturated, and conversely, below this range is undersaturated and above is oversaturated.
\end{itemize}

\noindent Moreover, to uncover the bulk region specifically, the following requirements are further imposed:
\begin{itemize}
    \item Bino component of the LSP must be greater than $99.9\%$ to prevent excessively large annihilation cross-sections from wino or Higgsino contributions, therefore, all bulk points will possess $N_{11} \simeq 1$ in Eq.~\ref{DMeq}.
    \item Condition $m_{h} \ll 2 m_{\Tilde{\chi}_{1}^{0}} \ll m_{H^{0}}$,$m_{A^{0}}$ is employed to avoid the Higgs funnel. In addition, when the condition $|\mu|^{2} \gg M_{Z}^{2}$ is satisfied, the SM Higgs resonance is significantly suppressed because its coupling is proportional to $g_{h{\Tilde{\chi}_{1}^{0}}{\Tilde{\chi}_{1}^{0}}} \propto \frac{M_{Z}(2\mu \text{cos}\beta+ M_{1})}{\mu^{2}-M_{1}^{2}}$~\cite{Djouadi:2005dz}. In the bulk region, $\mu > 1~\text{TeV} $, ensuring this condition is naturally satisfied.
    \item Choose $\mathcal{R}_{\Tilde{\tau}_{1}}\equiv \frac{m_{\Tilde{\tau}_{1}}-m_{\Tilde{\chi}_{1}^{0}}}{m_{\Tilde{\chi}_{1}^{0}}}  \gtrsim 10 \%$  and $\mathcal{R}_{\Tilde{e}_{R}}> \mathcal{R}_{\Tilde{\tau}_{1}}$ to  suppress coannihilation.
\end{itemize}

A light stau is not necessarily required for the bulk, though we do consider a light stau NLSP in this work since a right-handed smuon NLSP is subject to more strict collider constraints in Ref~\cite{ATLAS:2022hbt}. Nonetheless, the conclusions of our work pertaining to the bulk and natural DM do not change with a heavy light stau and smuon NLSP, though these details are not provided here given that our parameter space scans were focused on the light slepton region. It should be noted that the light stau NLSP we analyze here does indeed exceed the most recent and strongest light stau constraints of Ref.~\cite{ATLAS:2024fub}. It has also been assumed in the present work that $m_{\Tilde{e}_{R}} = m_{\Tilde{\mu}_{R}}$, but we shall break this degeneracy and thus present the scenario where only the LSP and right-handed smuon are light and all other sparticles are heavy in future publications.

All calculations are performed using the {\tt SuSpect3}~\cite{Djouadi:2002ze,Kneur:2022vwt} codebase to compute the supersymmetric particle mass spectrum and experimentally observable quantities, in unison with {\tt MicrOMEGAs 5.0}~\cite{Belanger:2018ccd} to calculate the DM relic density, including spin-dependent and spin-independent scattering cross-sections. The {\tt Suspect3} calculation of $\Delta a_{\mu}$ includes the leading two-loop QED correction, which is roughly a 7\% reduction of the one-loop contribution.

\begin{figure*}[htb]
  \centering
  \begin{minipage}{0.46\textwidth}
    \centering
    \includegraphics[width=\textwidth]{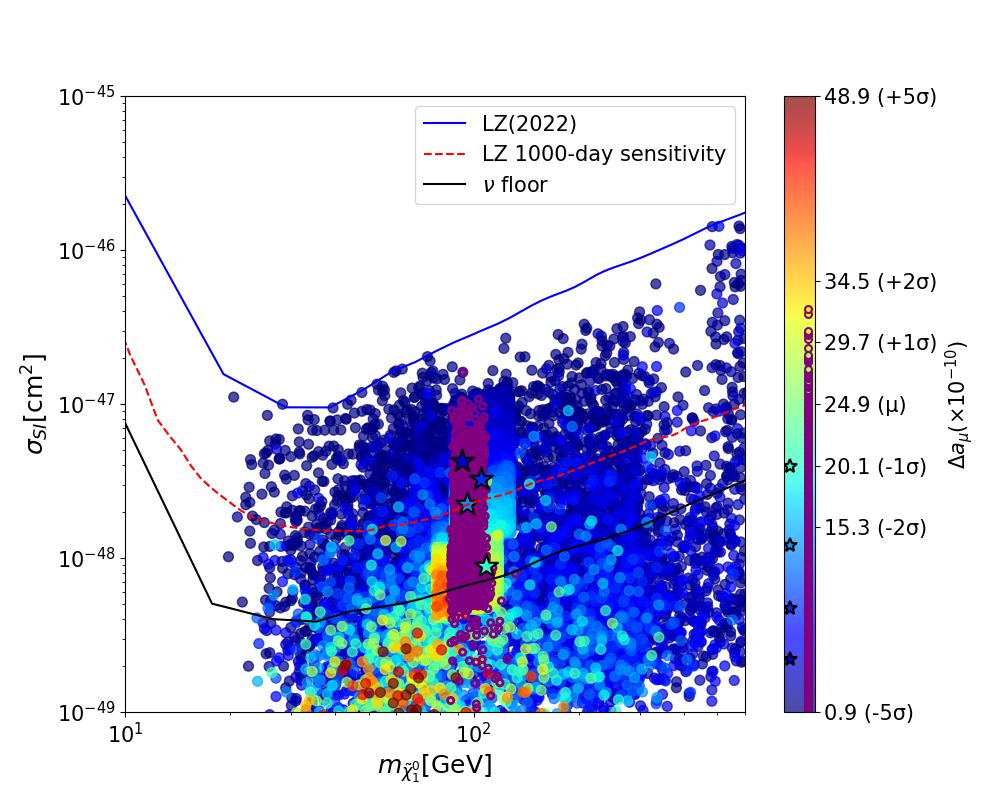}
    \par\smallskip
    \textbf{(a)}
  \end{minipage}
  \hspace{0.05\textwidth}
  \begin{minipage}{0.46\textwidth}
    \centering
    \includegraphics[width=\textwidth]{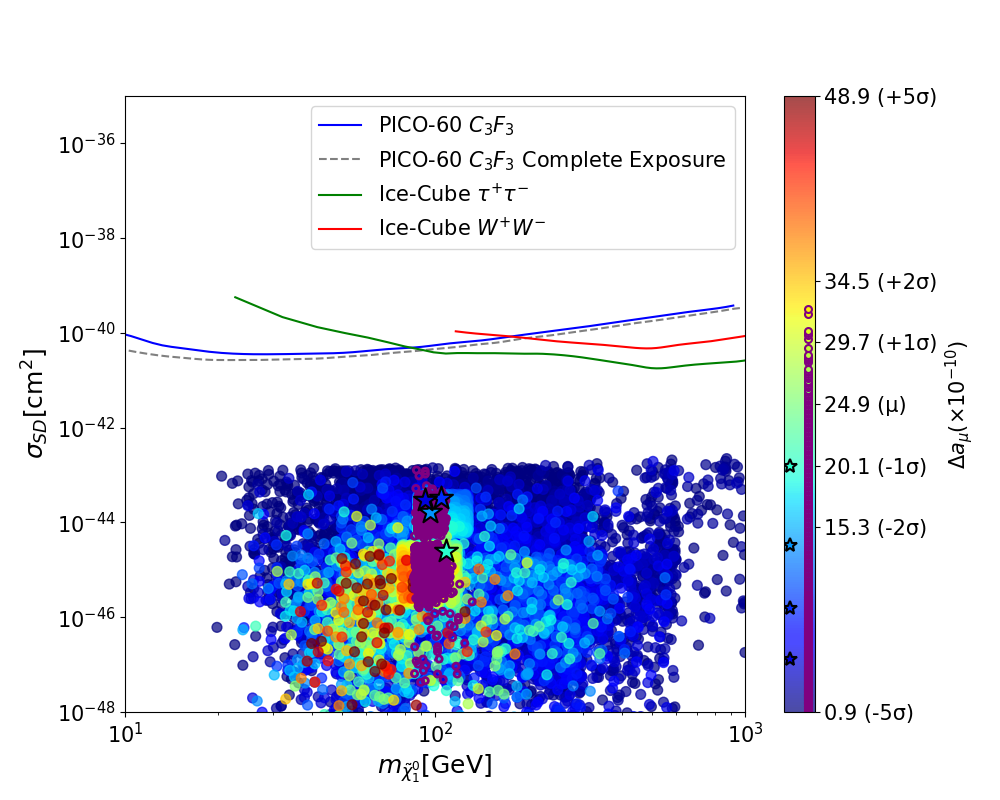}
    \par\smallskip
    \textbf{(b)}
  \end{minipage}
  \caption{Spin-independent DM-proton cross-sections versus $m_{\Tilde{\chi}_{1}^{0}}$ in the pMSSM are illustrated in the {\bf (a)} plot. The bulk rests beyond the constraints of the LZ 2022~\cite{LZ:2018qzl,LZ:2022lsv} version of the experiment, though the sensitivity of the future LZ 1000-day~\cite{LZ:2018qzl} experiment expects coverage of approximately half of this parameter space, including three benchmark points of Table~\ref{C4_T1_Benchmark}. In fact, the LZ 1000-day run will have the sensitivity to probe only up to $\Delta a_{\mu} \sim 14 \times 10^{-10}$. A minor subspace of the bulk is submerged below the neutrino floor~\cite{OHare:2021utq} where potential detection is hindered by large systematic uncertainties. The maximum $\Delta a_{\mu}$ in the bulk that may be probed by any direct-detection experiment without impact against the neutrino floor is $\Delta a_{\mu} \sim 30 \times 10^{-10}$, which is less than +2$\sigma$ above the BNL+FNAL central value of $\Delta a_{\mu} = 24.9 \times 10^{-10}$. The {\bf (b)} plot depicts the spin-dependent DM-proton cross-sections versus $m_{\Tilde{\chi}_{1}^{0}}$ in the pMSSM. The sensitivity of PICO-60 $C_{3}F_{3}$~\cite{PICO:2019vsc}, PICO-60 $C_{3}F_{3}$ Complete Exposure, and IceCube~\cite{IceCube:2016dgk} are highlighted. The same color scheme and constraints of Fig.~\ref{C4_F03_amuSmuonR} apply in this figure as well.}
  \label{C4_F09_chiSigmaSI}
\end{figure*}

\begin{figure}[ht]
    \centering
    \includegraphics[width=1.0\linewidth]{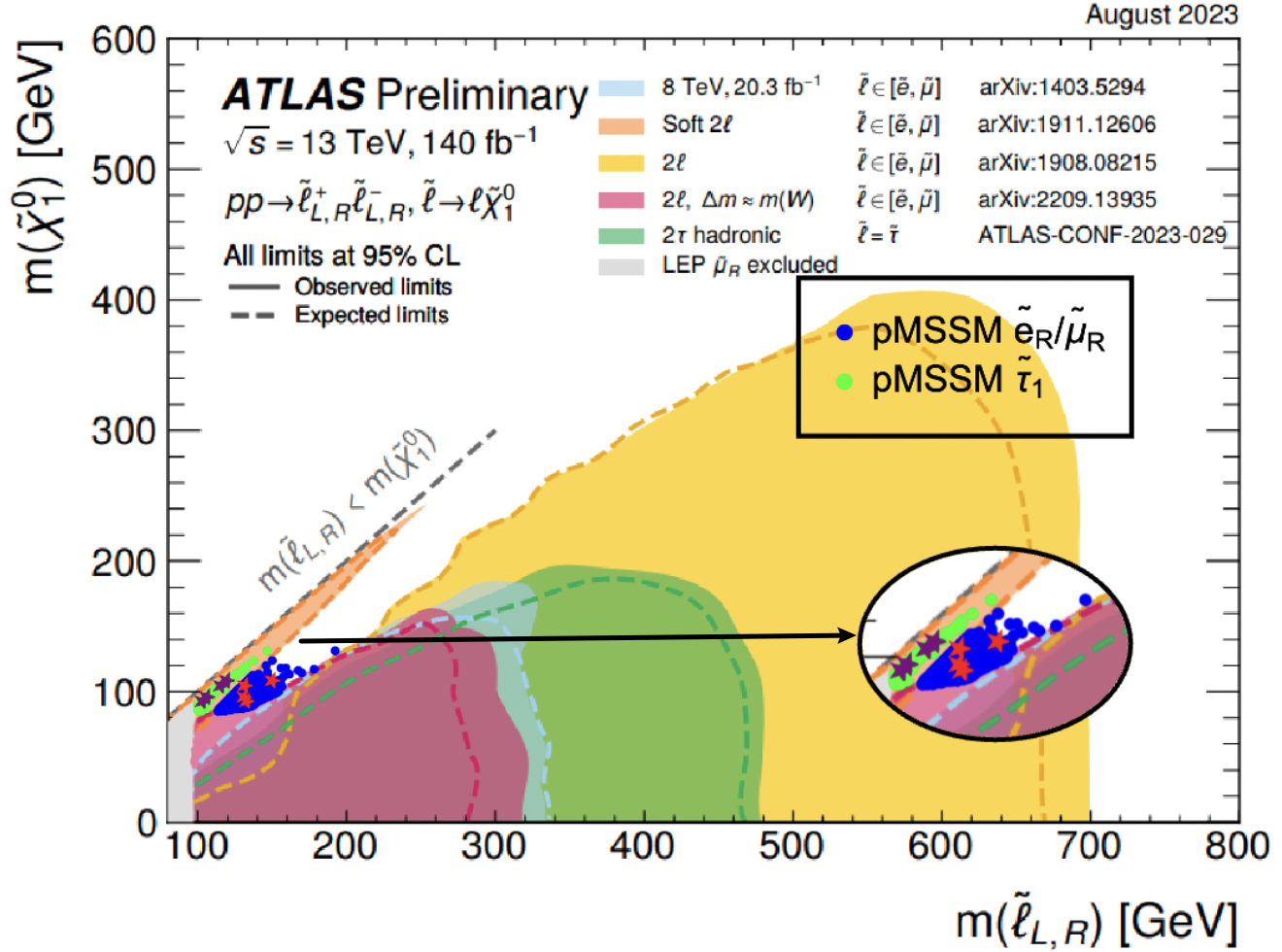}
    \caption{The pMSSM bulk regions superimposed over the ATLAS SUSY searches for electroweak production of sleptons~\cite{ATLAS:2023xco,ATLAS:2014zve,ATLAS:2019lng,ATLAS:2019lff,ATLAS:2022hbt,ATLAS:2023djh}. The blue and green points correspond to the $\Tilde{e}_{R}= \Tilde{\mu}_{R}$ and $\Tilde{\tau}_{1}$ in the pMSSM, respectively. Note that the ATLAS orange shaded sliver applies only to $\Tilde{e}_{R}= \Tilde{\mu}_{R}$, while the $\Tilde{\tau}_{1}$ constraints are shown in the green shaded region. The purple stars specifically refer to the light stau for each of the benchmarks in Table~\ref{C4_T1_Benchmark}, while the red stars refer to the first two generation right-handed sleptons $m_{\Tilde{e}_{R}} = m_{\Tilde{\mu}_{R}}$. The inset enlarges the bulk, which should remain a persistent priority of collider searches, both present and forthcoming.}
    \label{C4_F12_bulk}
\end{figure}

\section{Phenomenological Analysis}{\label{Sec4}}

The phenomenological analysis in this section will focus explicitly on the compressed mass region. The relationship between the bino mass $m_{\Tilde{\chi}^{0}_{1}}$ and $R_{\Tilde{\tau}_{1}}$ is illustrated in Fig.~\ref{C4_F01_bulk}. All points plot in Fig.~\ref{C4_F01_bulk} satisfy the experimental constraints outlined in the prior section. The magenta points represent the saturated DM relic density, while the cyan and gray points indicate undersaturated and oversaturated relic density, respectively. Given that an $\Tilde{e}_{R}$ NLSP is subject to more stringent collider constraints, we only study a $\Tilde{\tau}_{1}$ NLSP. The magenta region in Fig.~\ref{C4_F01_bulk} where $\mathcal{R}_{\Tilde{e}_{R}}> \mathcal{R}_{\Tilde{\tau}_{1}}\gtrsim 10\%$ is identified as the bulk, and as such can generate dark matter naturally. 

Several parameters critical to this pMSSM $g_{\mu}-2$ study are illustrated in Figs.~\ref{C4_F03_amuSmuonR}-\ref{C4_F09_chiSigmaSI}. A quick description of the plot theme in these figures is in order. On the right side of each plot is a color scale corresponding to a $\pm$5$\sigma$ range of $\Delta a_{\mu}$ around the BNL+FNAL central value of $\Delta a_\mu^{\text{BNL+FNAL}}= 24.9 \times 10^{-10}$, covering the total scope of BNL+FNAL, Lattice, and CMD-3 measured discrepancies from theoretical values. Every solid point drawn in the plot space is color coded to this $\Delta a_{\mu}$ color scale, providing rapid recognition of where the viable $g_{\mu}-2$ region stands within the larger model space. To enable identification of the bulk with ease, all bulk points are drawn as open purple circles (heavily overlapped in the denser areas), both in the plot space and color scale, highlighting the breadth of range of $\Delta a_{\mu}$ in the bulk. The stars with black outline denote four benchmark points for the bulk region, each corresponding to successively larger values of $\Delta a_{\mu}$. The specific values of all parameters of the corresponding marks are given in Table~\ref{C4_T1_Benchmark}, and the interior color of each star correlates to the precise color assigned to that particular $\Delta a_{\mu}$ value per the color scale. All points satisfy the experimental constraints listed in Sec.~\ref{Sec3} (first itemized subset of constraints), and the bulk points further comply with the bulk requirements shown in Sec.~\ref{Sec3} (second itemized subset of constraints).

To commence a survey of each pivotal parameter, the right-handed smuon and $m_{\chi^0_1}$ mass plane is presented in Fig.~\ref{C4_F03_amuSmuonR}{\bf (a)}. The bulk region covers a wide range of $\Delta a_{\mu}$, with the corresponding mass $m_{\tilde{\mu}_{R}}$ being within 100-200 GeV, as depicted by the open purple circles in Fig.~\ref{C4_F03_amuSmuonR}{\bf (a)}. However, all points that satisfy the experimental constraints of Sec.~\ref{Sec3} are shown in Fig.~\ref{C4_F03_amuSmuonR}{\bf (a)}. Those points with $\tilde{\mu}_R$ heavier than the bulk can generate the saturated relic density since the $\tilde{\tau}_1$ and $\tilde{e}_R$ are also light and can effectively reduce the relic density via annihilation with the LSP, though with coannihilation suppressed. Our intent in this work is to uncover the region where light right-handed smuons have evaded the LHC's reach thus far, so we are primarily interested in points in the lower left corner of Fig.~\ref{C4_F03_amuSmuonR}{\bf (a)} that would comprise the bulk and hence be subject to much more copious production at the LHC. Recall from Sec.~\ref{Sec1} that the recent BMW lattice calculations of the SM prediction of $a_{\mu}$ has shifted closer to the experimental measurement, yielding a smaller $\Delta a_{\mu}$, which is in fact consistent at 2$\sigma$ with no deviation at all between the SM theoretical value and the empirical result. Furthermore, the CMD-3 recalculations noted in Sec.~\ref{Sec1} are consistent with no deviation at 1$\sigma$. In the event the trajectory of $\Delta a_{\mu}$ continues toward a null value, it is important for the light right-handed smuons in the bulk to remain viable. The viability of the light right-handed smuons in the bulk is evident in Fig.~\ref{C4_F03_amuSmuonR}{\bf (a)} even if the BMW Lattice and CMD-3 results provide a more realistic description of $\Delta a_{\mu}$. The color scale just to the right of the plot space shows open purple circles that represent the bulk ranging from $0.9 \times 10^{-10} \lesssim \Delta a_{\mu} \lesssim 32 \times 10^{-10}$, comfortably consistent at 1$\sigma$ with each of the BNL+FNAL, BMW, and CMD-3 results. There are ample $m_{\tilde{\mu}_R} < 200$~GeV points in Fig.~\ref{C4_F03_amuSmuonR}{\bf (a)} with nearly null $\Delta a_{\mu}$, if by chance the SM theoretical calculations continue to close the gap with the BNL+FNAL experimental value. The relationship between the left-handed smuon mass $m_{\tilde{\mu}_{L}}$ and $m_{\chi^0_1}$ is shown in Fig.~\ref{C4_F03_amuSmuonR}{\bf (b)}. LHC SUSY search constraints necessitate a cut on the data points in Fig.~\ref{C4_F03_amuSmuonR}{\bf (b)} of $m_{\Tilde{\mu}_{L}} > 700$ GeV. It is apparent from Fig.~\ref{C4_F03_amuSmuonR}{\bf (b)} that within the bulk region, $m_{\tilde{\mu}_{L}}$ around 1 TeV can still cover a wide range of  $\Delta a_{\mu}$.

Larger values of $\tan \beta$ can enhance $a_{\mu}$ and produce larger discrepancies with the theoretically calculated value. This is evident in Fig.~\ref{C4_F05_amuTanb}, which delineates $\tan \beta$ as a function of $m_{\chi^0_1}$. Within the bulk region, $\tan \beta$ is in fact less than 40, as anticipated. The dependence of $\Delta a_{\mu}$ on the Gaugino mass parameter $M_2$ is not significant, and likewise, there is little correlation to the SUSY breaking $\mu$ term, light Higgs boson mass $m_h$, and $\Omega h^{2}$.

Concurrent with the ATLAS collaboration's attention to light right-handed smuon production in the coming years, nuclear recoil from elastic collisions between the bino LSP in the bulk and nucleons will also be under probe by direct detection experiments, both spin-independent and spin-dependent. The primary neutralino DM coupling to nuclei are scalar-mediated interactions via squarks and Higgs bosons, where the neutralino coupling to the nuclei is via the total mass of the nucleus of the target material in the spin-independent experiment. The neutralino DM coherently couples to the complete nucleus, therefore, the scattering cross-section is proportional to the mass-squared of the nucleus. This is independent of the spin and thus greatly enhances the cross-section with more massive nuclei such as xenon. On the contrary, alternating spins within nuclei suppress the spin-dependent cross-section, where the neutralino couples to the spin of the nucleus of the target material through mostly $Z^0$ exchange, with no enhancement from the nuclei mass. The resulting small event rate of a bino LSP in spin-dependent experiments means spin-dependent experiments have a much lower sensitivity to bino-nucleon scattering than spin-independent experiments. In addition, the spin-independent DM-proton and DM-neutron cross-sections vary only slightly due to comparable form-factors, hence we study the spin-independent DM-proton scattering cross-sections. Conversely, the form-factors differ more in spin-dependent elastic collisions, leading to a larger DM-proton cross-section than the DM-neutron cross-section, therefore, we study the spin-dependent DM-proton scattering cross-sections.

The direct detection prospects of the bulk in the pMSSM were first investigated in Ref.~\cite{Yin:2023jhs}. In tandem with the pMSSM analysis, it was observed that the bulk in No-Scale $\mathcal{F}$-$SU(5)$ should be thoroughly probed by the forthcoming LZ 1000-day experiment~\cite{LZ:2018qzl}. Conversely, Ref.~\cite{Yin:2023jhs} acknowledged that only approximately half of the bulk parameter space in the pMSSM could be reached by the 1000-day LZ run. To reveal this partial coverage of the pMSSM bulk by LZ in more detail, see Fig.~\ref{C4_F09_chiSigmaSI}{\bf (a)} expressing the bino mass $m_{\Tilde{\chi}_{1}^{0}}$ as a function of its spin-independent scattering cross-section with protons. Prominently featured in Fig.~\ref{C4_F09_chiSigmaSI}{\bf (a)} are constraints from LZ 2022~\cite{LZ:2018qzl,LZ:2022lsv}, the future LZ 1000-day experiment, and neutrino floor~\cite{OHare:2021utq}. Complicating a fine probe of the pMSSM bulk is the potential that a minor subspace is submerged below the neutrino floor, which carries large systematic uncertainties, though more optimistically, a large fraction exceeds the 1000-day LZ discovery threshold. Regarding the benchmarks of Table~\ref{C4_T1_Benchmark}, all four points comfortably eclipse the neutrino floor, but only three await conceivable discovery by the 1000-day LZ. Benchmark A corresponding to $\Delta a_{\mu}=20.01\times 10^{-10}$, visible in Figs.~\ref{C4_F03_amuSmuonR}-\ref{C4_F09_chiSigmaSI}, rests beyond the 1000-day LZ projected sensitivity, though on the contrary, the $\Delta a_{\mu}$ values for the remaining three points are handily accessible by the future LZ 1000-day experiment. This is true for the pMSSM bulk in general, where the LZ 1000-day run will have the sensitivity to probe only up to $\Delta a_{\mu} \sim 14 \times 10^{-10}$. The maximum $\Delta a_{\mu}$ in the bulk that may be probed by any direct-detection experiment without descending below the neutrino floor is about $\Delta a_{\mu} \sim 30 \times 10^{-10}$, which is less than +2$\sigma$ above the BNL+FNAL central value of $\Delta a_{\mu} = 24.9 \times 10^{-10}$. All pMSSM bulk points with $\Delta a_{\mu} \gtrsim 14 \times 10^{-10}$ will require a more sensitive direct-detection experiment.

As a complement to the spin-independent cross-sections, Fig.~\ref{C4_F09_chiSigmaSI}{\bf (b)} analyzes spin-dependent scattering cross-sections with protons. The chief experimental constraints delineated in Fig.~\ref{C4_F09_chiSigmaSI}{\bf (b)} are from PICO-60 $C_{3}F_{3}$, PICO-60 $C_{3}F_{3}$ Complete Exposure, and IceCube. The green and red lines in Fig.~\ref{C4_F09_chiSigmaSI}{\bf (b)} represent the constraints from IceCube on $\Tilde{\chi}_{1}^{0}$ annihilating into $\tau^{+}\tau^{-}$ and $W^{+}W^{-}$, respectively. The spin-dependent status emanating from Fig.~\ref{C4_F09_chiSigmaSI}{\bf (b)} is that the spin-dependent scattering cross-sections of DM lay well beyond the current experimental detection capabilities, posing a substantial challenge for future spin-dependent experiments. As anticipated for the scalar couplings of neutralino DM to target nuclei, the spin-dependent experiments in Fig.~\ref{C4_F09_chiSigmaSI}{\bf (b)} are much less sensitive to our bino LSP in the bulk than the spin-independent experiments in Fig.~\ref{C4_F09_chiSigmaSI}{\bf (a)}.

The paramount attribute of the bulk lies in its unsuccessful probing to date given its difficult to measure light compressed spectra. This fact is lucidly illuminated in Fig.~\ref{C4_F12_bulk}, where the bulk points are superimposed over the summary plot of ATLAS SUSY searches for electroweak production of sleptons~\cite{ATLAS:2023xco,ATLAS:2014zve,ATLAS:2019lng,ATLAS:2019lff,ATLAS:2022hbt,ATLAS:2023djh}. Although not reproduced in this paper, the present unprobed status of the bulk is analogous regarding CMS SUSY searches for electroweak production of sleptons~\cite{CMS:2020bfa,CMS:2023qhl,CMS:2022syk}. Notice that the orange region in Fig.~\ref{C4_F12_bulk} represents constraints on $\Tilde{e}_{R}$ rather than $\Tilde{\tau}_{1}$. The blue and green points correspond to the bulk region in the pMSSM, representing $\Tilde{e}_{R}$ and $\Tilde{\tau}_{1}$, respectively. The purple stars in Fig.~\ref{C4_F12_bulk} specifically refer to the light stau for each of the benchmarks in Table~\ref{C4_T1_Benchmark}, while the red stars refer to the first two generation right-handed sleptons $m_{\Tilde{e}_{R}} = m_{\Tilde{\mu}_{R}}$. It is obvious in Fig.~\ref{C4_F12_bulk} that the bulk has yet to be probed by the LHC. As was emphasized in Ref.~\cite{Yin:2023jhs}, the bulk can be tested at the FCC-ee and CEPC future circular colliders, though the more timely optimistic outcome would be observation at the LHC3, which ATLAS has confirmed as feasible.

\section{Conclusion}{\label{Sec5}}

The anomalous magnetic moment of the muon, commonly alluded to as $g_{\mu}-2$, swiftly displaced vintage topics in high-energy physics and escalated to the vanguard of particle physics dialog, and indeed, the scientific community at large. The rationale for the preponderance of discussion on the muon's magnetic moment derives from the 2021 BNL+FNAL announcement confirming the large measured $g_{\mu}-2$ deviation with theoretical predictions and substantially reduced uncertainty that could portend new physics. While the new physics was open to assorted theoretical explanations, supersymmetry stood as a conspicuous and popular solution to the physical dilemma. The leading-order supersymmetric contributions to muon $g_{\mu}-2$ mainly emanate from loops involving charginos/sneutrinos and neutralinos/smuons, obligating these supersymmetric particles to be relatively light. However, generating light neutralinos and sleptons without running afoul with contemporary LHC SUSY search contraints was more difficult to achieve in practice. For example, in the fully cMSSM, the steadily ascending lower mass limits on gluinos and squarks stalled all efforts of achieving very light neutralinos and smuons in that simplified model. The seemingly only way forward soon thereafter in 2021-2022 was to relax boundary conditions within the cMSSM framework to achieve larger contributions to muon $g_{\mu}-2$. Time would tell whether a more robust methodology would emerge that could surmount these problematic hurdles.

In 2023 the authors developed an elegant original approach to solving an age old puzzle regarding how dark matter can be generated naturally within the supersymmetric model space, without the need for fine-tuning the levers to produce a favorable match with relic density measurements. The method screened out finely tuned tactics such as coannihilation, Z/Higgs funnel, and mixing/well-tempered scenarios, and replaced them with a viable bulk region accommodating naturally light neutralinos and sleptons. In this initial study, both the physical model No-Scale ${\cal F}$-$SU(5)$ and the pMSSM model spaces were parsed for spectra that could generate the observed relic density via annihilation only, while concurrently conforming to all the latest experimentally derived data. That prior study was aimed primarily at uncovering the bulk in No-Scale ${\cal F}$-$SU(5)$, though it did provide a glimpse into the bulk within the pMSSM. This paper extends that original analysis and presents all the essential features of the bulk in the pMSSM and tangible evidence as to why to this day the bulk lingers unexplored at the LHC. 

The struggle the LHC has endured in effectively probing the bulk stems from the compressed nature of the light spectra. On a more optimistic note, ATLAS has confirmed that the bulk is within the reach of the LHC3, though the bulk will also certainly be accessible to the next generation circular machines, for instance, the FCC-ee and CEPC, anticipated to commence operations no sooner than the year 2040. Initial ingress into the bulk in the pMSSM is also expected by the upcoming LUX-ZEPLIN 1000-day experiment, given the manifestation of light neutralinos in the bulk, though the LZ 1000-day will only be able to probe about half the bulk's parameter space. Over and above the present unexplored state of the bulk, an interesting correlation occurred along the way to realizing naturally generated dark matter. In view of the compressed spectra and light neutralinos, the bulk also of course renders light sleptons, all less than about 200~GeV. This in consequence exposes that the bulk also supports a viable $g_{\mu}-2$, well within BNL+FNAL measurements, and the bulk has enough breadth to further satisfy CMD-3 results and discrepancies derived from lattice calculations. Experimental consistency does not end with $g_{\mu}-2$ either, where the bulk further accommodates constraints on all presently operating SUSY related experiments, including limits on sleptons, gluinos, squarks, rare B-meson decay, b-quark decays, light Higgs boson mass, and direct detection limits.

\begin{table*}
\large
\begin{tabular}{c|c|c|c|c}
  \hline
  & A (\begin{tikzpicture}
    \definecolor{Benchmark_A}{rgb}{0.16129, 1, 0.80645}
    \node[star, star points=5, star point ratio=2.25, minimum size=1cm,scale=0.5,fill=Benchmark_A, draw=black, line width=0.3mm] {};
\end{tikzpicture}) & B (\begin{tikzpicture}
    \definecolor{Benchmark_B}{rgb}{0, 0.58235, 1}
    \node[star, star points=5, star point ratio=2.25, minimum size=1cm,scale=0.5,fill=Benchmark_B, draw=black, line width=0.3mm] {};
\end{tikzpicture})& C (\begin{tikzpicture}
    \definecolor{Benchmark_C}{rgb}{0, 0.1745, 1.0}
    \node[star, star points=5, star point ratio=2.25, minimum size=1cm,scale=0.5,fill=Benchmark_C, draw=black, line width=0.3mm] {};
\end{tikzpicture}) & D (\begin{tikzpicture}
    \definecolor{Benchmark_D}{rgb}{0, 0, 0.87433}
    \node[star, star points=5, star point ratio=2.25, minimum size=1cm,scale=0.5,fill=Benchmark_D, draw=black, line width=0.3mm] {};
\end{tikzpicture})\\
  \hline
  \hline
     $M_{1}$& 111&98 & 107&95  \\
     $M_{2}$& 2026&3886 & 3644&3690  \\
     $M_{3}$& 3275&3691 & 3711&3798  \\
     $tan\beta$& 38.93& 38.20&34.18 &17.15  \\
     $M_{A}$& 1951& 2317& 2309&2302  \\
     $\mu$& 2326&1490 & 1332& 1327 \\
     $m_{\Tilde{q},{\Tilde{Q}}}$& 3762,4320&2626,2241 &2814,2344 &2616,2541  \\
     $m_{\Tilde{u}_{R},{\Tilde{d}_{R}}}$&3762,3762 &2626,2626 & 2814,2814& 2616,2616 \\
     $m_{\Tilde{t}_{R},{\Tilde{b}_{R}}}$&3840,2965 & 4520,3799& 4424,3968& 4487,3701  \\
     $m_{\Tilde{l},{\Tilde{L}}}$& 876,1912& 816,1718& 857,1988& 819,1991 \\
     $m_{\Tilde{e}_{R},{\Tilde{\tau}_{R}}}$&144,139 &124,114 &124,115 & 126,96 \\
     $A_{u,d}$&1045,3666 &1689,-4413 &1738,-4346 &1944,-4554  \\
     $A_{t,b}$&-2481,-1323 &-3884,-1422 &-3913,-1421 &-3838,-1595  \\
     $A_{e,\tau}$& 2386,1515& -347,-945& -387,-1022& -529,-966 \\
     \hline
     $m_{h}$& 122.1&123.4 &123.6 &123.3  \\
     $m_{H}$& 1951& 2317&2309 &2302  \\
     $m_{H^{\pm}}$& 1953&2319 &2311 &2304  \\
     \hline
     \colorbox{yellow}{$m_{\Tilde{\chi}_{1}^{0}}$},$m_{\Tilde{\chi}_{2}^{0}}$&\colorbox{yellow}{109},2047 &\colorbox{yellow}{96},1505 &\colorbox{yellow}{105},1348 &\colorbox{yellow}{93},1345   \\
     $m_{\Tilde{\chi}_{3,4}^{0}}$&2336,2346 & 1507,3827& 1350,3600&1347,3637 \\
     $m_{\Tilde{\chi}_{1,2}^{\pm}}$&2047,2346  & 1505,3827& 1348,3600&1345,3637 \\
     \hline
     $m_{\Tilde{g}}$&3568 &3747 &3784 &3853  \\
     $m_{\Tilde{u}_{L,R}}$&3874,3874 &2752,2752 &2938,2939 &2752,2752  \\
     $m_{\Tilde{t}_{1,2}}$& 3906,4413 &2392,4549 &2492,4455 &2697,4531\\
     $m_{\Tilde{d}_{L,R}}$&3848,3874 &2753,2752 &2939,2039 &2753,2753  \\
     $m_{\Tilde{b}_{1,2}}$&3107,4415 & 2375,3902& 2478,4067&2680,3816 \\
     $m_{\Tilde{e}_{L}} = m_{\Tilde{\mu}_{L}}$, $m_{\Tilde{e}_{R}} =$ \colorbox{Apricot}{$m_{\Tilde{\mu}_{R}}$}& 877,\colorbox{Apricot}{151}\begin{tikzpicture}
    \definecolor{Benchmark_E}{rgb}{1.0, 0, 0}
    \node[star, star points=5, star point ratio=2.25, minimum size=1cm,scale=0.5,fill=Benchmark_E]{};
\end{tikzpicture}& 817,\colorbox{Apricot}{132}\begin{tikzpicture}
    \definecolor{Benchmark_F}{rgb}{1.0, 0, 0}
    \node[star, star points=5, star point ratio=2.25, minimum size=1cm,scale=0.5,fill=Benchmark_F]{};
\end{tikzpicture}& 858,\colorbox{Apricot}{132}\begin{tikzpicture}
    \definecolor{Benchmark_G}{rgb}{1.0, 0, 0}
    \node[star, star points=5, star point ratio=2.25, minimum size=1cm,scale=0.5,fill=Benchmark_G]{};
\end{tikzpicture}& 820,\colorbox{Apricot}{133}\begin{tikzpicture}
    \definecolor{Benchmark_H}{rgb}{1.0, 0, 0}
    \node[star, star points=5, star point ratio=2.25, minimum size=1cm,scale=0.5,fill=Benchmark_H]{};
\end{tikzpicture}\\
     \colorbox{SpringGreen}{$m_{\Tilde{\tau}_{1}}$},$m_{\Tilde{\tau}_{2}}$&\colorbox{SpringGreen}{120}\begin{tikzpicture}
    \definecolor{Benchmark_I}{rgb}{0.5, 0, 0.5}
    \node[star, star points=5, star point ratio=2.25, minimum size=1cm,scale=0.5,fill=Benchmark_I]{};
\end{tikzpicture},1914 & \colorbox{SpringGreen}{107}\begin{tikzpicture}
    \definecolor{Benchmark_J}{rgb}{0.5, 0, 0.5}
    \node[star, star points=5, star point ratio=2.25, minimum size=1cm,scale=0.5,fill=Benchmark_J]{};
\end{tikzpicture},1720&\colorbox{SpringGreen}{116}\begin{tikzpicture}
    \definecolor{Benchmark_K}{rgb}{0.5, 0, 0.5}
    \node[star, star points=5, star point ratio=2.25, minimum size=1cm,scale=0.5,fill=Benchmark_K]{};
\end{tikzpicture},1989 & \colorbox{SpringGreen}{104}\begin{tikzpicture}
    \definecolor{Benchmark_L}{rgb}{0.5, 0, 0.5}
    \node[star, star points=5, star point ratio=2.25, minimum size=1cm,scale=0.5,fill=Benchmark_L]{};
\end{tikzpicture},1992 \\
     \hline
      $N_{11}$& 0.9998&0.9995&0.9994 &0.9994 \\
      $N_{12}$&$-8\times 10^{-5}$ &$-7\times 10^{-5}$ &$-1\times 10^{-4}$ &$-1\times 10^{-4}$ \\
      $N_{13}$&0.019 & 0.030& 0.034&0.034 \\
      $N_{14}$&-0.001 &-0.003& -0.004&-0.004\\
     \hline
     $\text{BR}(B_s^0\rightarrow \mu^+\mu^-)\times10^{-9}$&3.13 &3.23 &3.15 & 3.04  \\
     $\text{BR}(b\rightarrow s \gamma)\times10^{-4}$& 3.31& 3.24& 3.24& 3.31 \\
     $\Delta a_{\mu}\times 10^{-10}$& 20.1 & 13.9 & 9.0& 5.0\\
     $\Omega_{\Tilde{\chi}}h^{2}$&0.125 &0.116 &0.125 &0.120 \\
     $\sigma_{SI}\times10^{-12}[\text{pb}]$& 0.89& 2.2&3.3&4.3 \\
     $\sigma_{SD}\times10^{-8}[\text{pb}]$& 0.25&1.7 & 3.3&3.0\\
     \hline
     \end{tabular}     
    \caption{Four benchmark points residing in the bulk in the pMSSM. All masses are in GeV. The A, B, C, D stars with the black outline in the top row identify the precise location in the parameter space in Figs.~\ref{C4_F03_amuSmuonR}-\ref{C4_F09_chiSigmaSI} for each of the four benchmarks, with color matched to the numerical $\Delta a_{\mu}$ color scale just to the right of the plot space in Figs.~\ref{C4_F03_amuSmuonR}-\ref{C4_F09_chiSigmaSI}. The purple stars in the lower half of this Table specifically refer to the light stau for each of the benchmarks in the ATLAS slepton search Fig.~\ref{C4_F12_bulk}, while in Fig.~\ref{C4_F12_bulk} the red stars refer to the first two generation right-handed sleptons $m_{\Tilde{e}_{R}} = m_{\Tilde{\mu}_{R}}$. The highlights in this Table bring the reader's prompt attention to the LSP, right-handed smuon, and light stau masses.}
    \label{C4_T1_Benchmark}
\end{table*}

The vacant cupboard of substantiated SUSY events at the LHC has not deterred those of us who recognize SUSY as an indispensable field theory crucial for constructing high-energy fundamental physics and a Theory of Everything (TOE). Meanwhile, the LHC proton-proton collisions forge on and the floor on sparticle masses levitates higher and higher, which at worst could eventually deny SUSY of one of its most appealing attributes, that is, solving the hierarchy problem, or at best provoke severe tension as a solution to the problem. Supersymmetry could yet be empirically revealed in a spectrum with greater than 1 TeV neutralinos and sleptons, as only a cursory review of the current broad LHC constraints might informally suggest. At this belated stage though, the less than 200~GeV neutralino and slepton regions such as the bulk may offer a much more theoretically motivated resolution intimately connected with naturalness. Therefore, we suggest the LHC3 and HL-LHC detour from the heavy slepton realm and attempt to probe an obscure region such as the bulk where SUSY and dark matter could still be lurking and patiently awaiting discovery.

\begin{acknowledgments}
     T.L. would like to thank Xuai Zhuang for helpful discussions. We thank the ChinaHPC for providing us with the computational platform. This research is supported in part by the National Key Research and Development Program of China Grant No. 2020YFC2201504, by the Projects No. 11875062, No. 11947302, No. 12047503, No. 12275333, and No. 12347101 supported by the National Natural Science Foundation of China, by the Key Research Program of the Chinese Academy of Sciences, Grant No. XDPB15, by the Scientific Instrument Developing Project of the Chinese Academy of Sciences, Grant No. YJKYYQ20190049, by the International Partnership Program of Chinese Academy of Sciences for Grand Challenges, Grant No. 112311KYSB20210012, and by DOE Grant No. DE-FG02-13ER42020 (DVN).

\end{acknowledgments}




\begin{thebibliography}{}

\bibitem{Nilles:1983ge}
H.~P.~Nilles,
\href{https://www.sciencedirect.com/science/article/abs/pii/0370157384900085?via%3Dihub}{Phys. Rept. \textbf{110}, 1-162 (1984)}
\bibitem{Haber:1984rc}
H.~E.~Haber and G.~L.~Kane,
\href{https://www.sciencedirect.com/science/article/abs/pii/0370157385900511?via%3Dihub}{Phys. Rept. \textbf{117}, 75-263 (1985)}

\bibitem{MSSMWorkingGroup:1998fiq}
A.~Djouadi \textit{et al.} [MSSM Working Group],
[\href{https://arxiv.org/abs/hep-ph/9901246}{arXiv:hep-ph/9901246 [hep-ph]}].

\bibitem{Kane:1993td}
G.~L.~Kane, C.~F.~Kolda, L.~Roszkowski and J.~D.~Wells,
Phys. Rev. D \textbf{49}, 6173-6210 (1994)
[\href{https://arxiv.org/abs/hep-ph/9312272}{arXiv:hep-ph/9312272 [hep-ph]}].
J.~R.~Ellis, T.~Falk, G.~Ganis, K.~A.~Olive and M.~Srednicki,
Phys. Lett. B \textbf{510}, 236-246 (2001)
[\href{https://arxiv.org/abs/hep-ph/0102098}{arXiv:hep-ph/0102098 [hep-ph]}].
J.~R.~Ellis, K.~A.~Olive and Y.~Santoso,
New J. Phys. \textbf{4}, 32 (2002)
[\href{https://arxiv.org/abs/hep-ph/0202110}{arXiv:hep-ph/0202110 [hep-ph]}].
H.~Baer, C.~Balazs, A.~Belyaev, J.~K.~Mizukoshi, X.~Tata and Y.~Wang,
JHEP \textbf{07}, 050 (2002)
[\href{https://arxiv.org/abs/hep-ph/0205325}{arXiv:hep-ph/0205325 [hep-ph]}].
S.~S.~AbdusSalam, B.~C.~Allanach, H.~K.~Dreiner, J.~Ellis, U.~Ellwanger, J.~Gunion, S.~Heinemeyer, M.~Kraemer, M.~L.~Mangano and K.~A.~Olive, \textit{et al.}
Eur. Phys. J. C \textbf{71}, 1835 (2011)
[\href{https://arxiv.org/abs/1109.3859}{arXiv:1109.3859 [hep-ph]}].
M.~Chakraborti, L.~Roszkowski and S.~Trojanowski,
JHEP \textbf{05}, 252 (2021)
[\href{https://arxiv.org/abs/2104.04458}{arXiv:2104.04458 [hep-ph]}].
Z.~Li, G.~L.~Liu, F.~Wang, J.~M.~Yang and Y.~Zhang,
JHEP \textbf{12}, 219 (2021)
[\href{https://arxiv.org/abs/2106.04466}{arXiv:2106.04466 [hep-ph]}].

\bibitem{Ellis:1984bm}
J.~R.~Ellis, C.~Kounnas and D.~V.~Nanopoulos,
\href{https://www.sciencedirect.com/science/article/abs/pii/0550321384905558?via%3Dihub}{Nucl. Phys. B \textbf{247}, 373-395 (1984)}
M.~Drees,
\href{https://www.sciencedirect.com/science/article/abs/pii/0370269385904423?via%3Dihub}{Phys. Lett. B \textbf{158}, 409-412 (1985)}
J.~R.~Ellis, K.~Enqvist, D.~V.~Nanopoulos and K.~Tamvakis,
\href{https://www.sciencedirect.com/science/article/abs/pii/0370269385915916?via%3Dihub}{Phys. Lett. B \textbf{155}, 381-386 (1985)}
M.~Drees,
\href{https://journals.aps.org/prd/abstract/10.1103/PhysRevD.33.1468}{Phys. Rev. D \textbf{33}, 1468 (1986)}

\bibitem{Wang:2021bcx}
F.~Wang, L.~Wu, Y.~Xiao, J.~M.~Yang and Y.~Zhang,
Nucl. Phys. B \textbf{970}, 115486 (2021)
[\href{https://arxiv.org/abs/2104.03262}{arXiv:2104.03262 [hep-ph]}].
\bibitem{Wang:2018vrr}
F.~Wang, K.~Wang, J.~M.~Yang and J.~Zhu,
JHEP \textbf{12}, 041 (2018)
[\href{https://arxiv.org/abs/1808.10851}{arXiv:1808.10851 [hep-ph]}].
\bibitem{Aboubrahim:2021xfi}
A.~Aboubrahim, M.~Klasen and P.~Nath,
Phys. Rev. D \textbf{104}, no.3, 035039 (2021)
[\href{https://arxiv.org/abs/2104.03839}{arXiv:2104.03839 [hep-ph]}].

\bibitem{Baer:2004fu}
H.~Baer, A.~Mustafayev, S.~Profumo, A.~Belyaev and X.~Tata,
Phys. Rev. D \textbf{71}, 095008 (2005)
[\href{https://arxiv.org/abs/hep-ph/0412059}{arXiv:hep-ph/0412059 [hep-ph]}].

\bibitem{Baer:2005bu}
H.~Baer, A.~Mustafayev, S.~Profumo, A.~Belyaev and X.~Tata,
JHEP \textbf{07}, 065 (2005)
[\href{https://arxiv.org/abs/hep-ph/0504001}{arXiv:hep-ph/0504001 [hep-ph]}].
\bibitem{Ellis:2008eu}
J.~R.~Ellis, K.~A.~Olive and P.~Sandick,
Phys. Rev. D \textbf{78}, 075012 (2008)
[\href{https://arxiv.org/abs/0805.2343}{arXiv:0805.2343 [hep-ph]}].
\bibitem{Ellis:2012nv}
J.~Ellis, F.~Luo, K.~A.~Olive and P.~Sandick,
Eur. Phys. J. C \textbf{73}, no.4, 2403 (2013)
[\href{https://arxiv.org/abs/1212.4476}{arXiv:1212.4476 [hep-ph]}].
\bibitem{Buchmueller:2013rsa}
O.~Buchmueller, R.~Cavanaugh, A.~De Roeck, M.~J.~Dolan, J.~R.~Ellis, H.~Flacher, S.~Heinemeyer, G.~Isidori, J.~Marrouche and D.~Martinez Santos, \textit{et al.}
Eur. Phys. J. C \textbf{74}, no.6, 2922 (2014)
[\href{https://arxiv.org/abs/1312.5250}{arXiv:1312.5250 [hep-ph]}].

\bibitem{Ellis:2002wv}
J.~R.~Ellis, K.~A.~Olive and Y.~Santoso,
Phys. Lett. B \textbf{539}, 107-118 (2002)
[\href{https://arxiv.org/abs/hep-ph/0204192}{arXiv:hep-ph/0204192 [hep-ph]}].
\bibitem{Ellis:2002iu}
J.~R.~Ellis, T.~Falk, K.~A.~Olive and Y.~Santoso,
Nucl. Phys. B \textbf{652}, 259-347 (2003)
[\href{https://arxiv.org/abs/hep-ph/0210205}{arXiv:hep-ph/0210205 [hep-ph]}].
\bibitem{Buchmueller:2014yva}
O.~Buchmueller, R.~Cavanaugh, M.~Citron, A.~De Roeck, M.~J.~Dolan, J.~R.~Ellis, H.~Flaecher, S.~Heinemeyer, S.~Malik and J.~Marrouche, \textit{et al.}
Eur. Phys. J. C \textbf{74}, no.12, 3212 (2014)
[\href{https://arxiv.org/abs/1408.4060}{arXiv:1408.4060 [hep-ph]}].

\bibitem{Baer:2021aax}
H.~Baer, V.~Barger and H.~Serce,
Phys. Lett. B \textbf{820}, 136480 (2021)
[\href{https://arxiv.org/abs/2104.07597}{arXiv:2104.07597 [hep-ph]}].

\bibitem{Ellis:2024ijt}
J.~Ellis, K.~A.~Olive and V.~C.~Spanos,
[\href{https://arxiv.org/abs/2407.08679}{arXiv:2407.08679 [hep-ph]}].

\bibitem{Muong-2:2021ojo}
B.~Abi \textit{et al.} [Muon g-2],
Phys. Rev. Lett. \textbf{126}, no.14, 141801 (2021)
[\href{https://arxiv.org/pdf/2104.03281}{arXiv:2104.03281 [hep-ex]}].

\bibitem{Muong-2:2023cdq}
D.~P.~Aguillard \textit{et al.} [Muon g-2],
Phys. Rev. Lett. \textbf{131}, no.16, 161802 (2023)
[\href{https://arxiv.org/pdf/2308.06230}{arXiv:2308.06230 [hep-ex]}].

\bibitem{Borsanyi:2020mff}
S.~Borsanyi, Z.~Fodor, J.~N.~Guenther, C.~Hoelbling, S.~D.~Katz, L.~Lellouch, T.~Lippert, K.~Miura, L.~Parato and K.~K.~Szabo, \textit{et al.}
Nature \textbf{593}, no.7857, 51-55 (2021)
[\href{https://arxiv.org/pdf/2002.12347}{arXiv:2002.12347 [hep-lat]}].

\bibitem{Kuberski:2023qgx}
S.~Kuberski,
PoS \textbf{LATTICE2023}, 125 (2024)
[\href{https://arxiv.org/pdf/2312.13753}{arXiv:2312.13753 [hep-lat]}].

\bibitem{Yin:2023jhs}
X.~Yin, J.~A.~Maxin, D.~V.~Nanopoulos and T.~Li,
Phys. Rev. D \textbf{109}, no.11, 115031 (2024)
[\href{https://arxiv.org/abs/2310.03622}{arXiv:2310.03622 [hep-ph]}].

\bibitem{Hooper:2002nq}
D.~Hooper and T.~Plehn,
Phys. Lett. B \textbf{562}, 18-27 (2003)
[\href{https://arxiv.org/abs/hep-ph/0212226}{arXiv:0212226 [hep-ph]}].
A.~Bottino, N.~Fornengo and S.~Scopel,
Phys. Rev. D \textbf{67}, 063519 (2003)
[\href{https://arxiv.org/abs/hep-ph/0212379}{arXiv:0212379 [hep-ph]}].
H.~K.~Dreiner, S.~Heinemeyer, O.~Kittel, U.~Langenfeld, A.~M.~Weber and G.~Weiglein,
Eur. Phys. J. C \textbf{62}, 547-572 (2009)
[\href{https://arxiv.org/abs/0901.3485}{arXiv:0901.3485 [hep-ph]}].
D.~Feldman, Z.~Liu and P.~Nath,
Phys. Rev. D \textbf{81}, 117701 (2010)
[\href{https://arxiv.org/abs/1003.0437}{arXiv:1003.0437 [hep-ph]}].
D.~Albornoz Vasquez, G.~Belanger, C.~Boehm, A.~Pukhov and J.~Silk,
Phys. Rev. D \textbf{82}, 115027 (2010)
[\href{https://arxiv.org/abs/1009.4380}{arXiv:1009.4380 [hep-ph]}].
D.~Albornoz Vasquez, G.~Belanger and C.~Boehm,
Phys. Rev. D \textbf{84}, 095015 (2011)
[\href{https://arxiv.org/abs/1108.1338}{arXiv:1108.1338 [hep-ph]}].
A.~Arbey, M.~Battaglia and F.~Mahmoudi,
Eur. Phys. J. C \textbf{72}, 2169 (2012)
[\href{https://arxiv.org/abs/1205.2557}{arXiv:1205.2557 [hep-ph]}].
P.~Grothaus, M.~Lindner and Y.~Takanishi,
JHEP \textbf{07}, 094 (2013)
[\href{https://arxiv.org/abs/1207.4434}{arXiv:1207.4434 [hep-ph]}].
C.~Boehm, P.~S.~B.~Dev, A.~Mazumdar and E.~Pukartas,
JHEP \textbf{06}, 113 (2013)
[\href{https://arxiv.org/abs/1303.5386}{arXiv:1303.5386 [hep-ph]}].
A.~Arbey, M.~Battaglia and F.~Mahmoudi,
Phys. Rev. D \textbf{88}, 095001 (2013)
[\href{https://arxiv.org/abs/1308.2153}{arXiv:1308.2153 [hep-ph]}].
L.~Calibbi, J.~M.~Lindert, T.~Ota and Y.~Takanishi,
JHEP \textbf{11}, 106 (2014)
[\href{https://arxiv.org/abs/1410.5730}{arXiv:1410.5730 [hep-ph]}].
J.~Cao, Y.~He, L.~Shang, W.~Su and Y.~Zhang,
JHEP \textbf{03}, 207 (2016)
[\href{https://arxiv.org/abs/1511.05386}{arXiv:1511.05386 [hep-ph]}].
S.~Profumo and T.~Stefaniak,
Phys. Rev. D \textbf{94}, no.9, 095020 (2016)
[\href{https://arxiv.org/abs/1608.06945}{arXiv:1608.06945 [hep-ph]}].
R.~K.~Barman, G.~Belanger, B.~Bhattacherjee, R.~Godbole, G.~Mendiratta and D.~Sengupta,
Phys. Rev. D \textbf{95}, no.9, 095018 (2017)
[\href{https://arxiv.org/abs/1703.03838}{arXiv:1703.03838 [hep-ph]}].
F.~Ambrogi, S.~Kraml, S.~Kulkarni, U.~Laa, A.~Lessa and W.~Waltenberger,
Eur. Phys. J. C \textbf{78}, no.3, 215 (2018)
[\href{https://arxiv.org/abs/1707.09036}{arXiv:1707.09036 [hep-ph]}].
G.~Pozzo and Y.~Zhang,
Phys. Lett. B \textbf{789}, 582-591 (2019)
[\href{https://arxiv.org/abs/1807.01476}{arXiv:1807.01476 [hep-ph]}].
R.~Kumar Barman, G.~Belanger and R.~M.~Godbole,
Eur. Phys. J. ST \textbf{229}, no.21, 3159-3185 (2020)
[\href{https://arxiv.org/abs/2010.11674}{arXiv:2010.11674 [hep-ph]}].
M.~Van Beekveld, W.~Beenakker, M.~Schutten and J.~De Wit,
SciPost Phys. \textbf{11}, no.3, 049 (2021)
[\href{https://arxiv.org/abs/2104.03245}{arXiv:2104.03245 [hep-ph]}].
R.~K.~Barman, G.~B\'elanger, B.~Bhattacherjee, R.~M.~Godbole and R.~Sengupta,
Phys. Rev. Lett. \textbf{131}, no.1, 011802 (2023)
[\href{https://arxiv.org/abs/2207.06238}{arXiv:2207.06238 [hep-ph]}].
G.~Aad \textit{et al.} [ATLAS],
JHEP \textbf{05}, 106 (2024)
[\href{https://arxiv.org/abs/2402.01392}{arXiv:2402.01392 [hep-ex]}].

\bibitem{Ellis:1983ew}
J.~R.~Ellis, J.~S.~Hagelin, D.~V.~Nanopoulos, K.~A.~Olive and M.~Srednicki,
\href{https://www.sciencedirect.com/science/article/abs/pii/0550321384904619?via\%3Dihub}{Nucl. Phys. B \textbf{238} (1984), 453-476}

\bibitem{fsu5old}
S.~M.~Barr,
\href{https://www.sciencedirect.com/science/article/abs/pii/0370269382909662?via\%3Dihub}{Phys. Lett. B \textbf{112}, 219-222 (1982)}.
S.~M.~Barr,
\href{https://journals.aps.org/prd/abstract/10.1103/PhysRevD.40.2457}{Phys. Rev. D \textbf{40}, 2457 (1989)}.
J.~P.~Derendinger, J.~E.~Kim and D.~V.~Nanopoulos,
\href{https://www.sciencedirect.com/science/article/abs/pii/0370269384912383?via\%3Dihub}{Phys. Lett. B \textbf{139}, 170-176 (1984)}.
I.~Antoniadis, J.~R.~Ellis, J.~S.~Hagelin and D.~V.~Nanopoulos,
\href{https://www.sciencedirect.com/science/article/abs/pii/0370269387905338?via\%3Dihub}{Phys. Lett. B \textbf{194}, 231-235 (1987)};
\href{https://www.sciencedirect.com/science/article/abs/pii/0370269388909781?via\%3Dihub}{Phys. Lett. B \textbf{205}, 459-465 (1988)};
\href{https://www.sciencedirect.com/science/article/abs/pii/0370269388904194?via\%3Dihub}{Phys. Lett. B \textbf{208}, 209-215 (1988), Phys.Lett.B 213 (1988) 562 (addendum)};
\href{https://www.sciencedirect.com/science/article/abs/pii/0370269389901159?via\%3Dihub}{Phys. Lett. B \textbf{231}, 65-74 (1989)}.

\bibitem{fsu5}
T.~Li, J.~A.~Maxin, D.~V.~Nanopoulos and J.~W.~Walker,
Phys. Rev. D \textbf{83} (2011), 056015
[\href{https://arxiv.org/abs/1007.5100}{arXiv:1007.5100 [hep-ph]}].
T.~Li, J.~A.~Maxin, D.~V.~Nanopoulos and J.~W.~Walker,
Phys. Lett. B \textbf{710} (2012), 207-214
[\href{https://arxiv.org/abs/1112.3024}{arXiv:1112.3024 [hep-ph]}].

\bibitem{LZ:2018qzl}
D.~S.~Akerib \textit{et al.} [LZ],
Phys. Rev. D \textbf{101}, no.5, 052002 (2020)
[\href{https://arxiv.org/abs/1802.06039}{arXiv:1802.06039 [astro-ph.IM]}].

\bibitem{LZ:2022lsv}
J.~Aalbers \textit{et al.} [LZ],
Phys. Rev. Lett. \textbf{131}, no.4, 041002 (2023)
[\href{https://arxiv.org/abs/2207.03764}{arXiv:2207.03764 [hep-ex]}].

\bibitem{FCC:2018byv}
A.~Abada \textit{et al.} [FCC],
\href{https://link.springer.com/article/10.1140/epjc/s10052-019-6904-3}{Eur. Phys. J. C \textbf{79}, no.6, 474 (2019)}

\bibitem{FCC:2018evy}
A.~Abada \textit{et al.} [FCC],
\href{https://link.springer.com/article/10.1140/epjst/e2019-900045-4}{Eur. Phys. J. ST \textbf{228}, no.2, 261-623 (2019)}

\bibitem{CEPCStudyGroup:2018ghi}
J.~B.~Guimar\~aes da Costa \textit{et al.} [CEPC Study Group],
[\href{https://arxiv.org/abs/1811.10545}{arXiv:1811.10545 [hep-ex]}].

\bibitem{CMD-3:2023rfe}
F.~V.~Ignatov \textit{et al.} [CMD-3],
Phys. Rev. Lett. \textbf{132}, no.23, 231903 (2024)
[\href{https://arxiv.org/abs/2309.12910}{arXiv:2309.12910 [hep-ex]}].

\bibitem{Davier:2023fpl}
M.~Davier, A.~Hoecker, A.~M.~Lutz, B.~Malaescu and Z.~Zhang,
Eur. Phys. J. C \textbf{84}, no.7, 721 (2024)
[\href{https://arxiv.org/abs/2312.02053}{arXiv:2312.02053 [hep-ph]}].

\bibitem{Martin:1997ns}
S.~P.~Martin,
Adv. Ser. Direct. High Energy Phys. \textbf{18}, 1-98 (1998)
[\href{https://arxiv.org/abs/hep-ph/9709356}{arXiv:hep-ph/9709356 [hep-ph]}].

\bibitem{Arkani-Hamed:2006wnf}
N.~Arkani-Hamed, A.~Delgado and G.~F.~Giudice,
Nucl. Phys. B \textbf{741}, 108-130 (2006)
[\href{https://arxiv.org/abs/hep-ph/0601041}{arXiv:hep-ph/0601041 [hep-ph]}].

\bibitem{Lopez:1993vi}
J.~L.~Lopez, D.~V.~Nanopoulos and X.~Wang,
Phys. Rev. D \textbf{49} (1994), 366-372
[\href{https://arxiv.org/abs/hep-ph/9308336}{arXiv:hep-ph/9308336 [hep-ph]}].

\bibitem{Moroi:1995yh}
T.~Moroi,
Phys. Rev. D \textbf{53}, 6565-6575 (1996)
[erratum: Phys. Rev. D \textbf{56}, 4424 (1997)]
[\href{https://arxiv.org/abs/hep-ph/9512396}{arXiv:hep-ph/9512396 [hep-ph]}].
\bibitem{Martin:2001st}
S.~P.~Martin and J.~D.~Wells,
Phys. Rev. D \textbf{64}, 035003 (2001)
[\href{https://arxiv.org/abs/hep-ph/0103067}{arXiv:hep-ph/0103067 [hep-ph]}].

\bibitem{ATLAS:2019lff}
G.~Aad \textit{et al.} [ATLAS],
Eur. Phys. J. C \textbf{80}, no.2, 123 (2020)
[\href{https://arxiv.org/abs/1908.08215}{arXiv:1908.08215 [hep-ex]}].

\bibitem{ATLAS:2017mjy}
M.~Aaboud \textit{et al.} [ATLAS],
Phys. Rev. D \textbf{97}, no.11, 112001 (2018)
[\href{https://arxiv.org/abs/1712.02332}{arXiv:1712.02332 [hep-ex]}].

\bibitem{Vami:2019slp}
T.~A.~Vami [ATLAS and CMS],
PoS \textbf{LHCP2019}, 168 (2019)
[\href{https://arxiv.org/abs/1909.11753}{arXiv:1909.11753 [hep-ex]}].

\bibitem{CMS:2017okm}
A.~M.~Sirunyan \textit{et al.} [CMS],
Eur. Phys. J. C \textbf{77}, no.10, 710 (2017)
[\href{https://arxiv.org/abs/1705.04650}{arXiv:1705.04650 [hep-ex]}].

\bibitem{CMS:2014xfa}
V.~Khachatryan \textit{et al.} [CMS and LHCb],
Nature \textbf{522}, 68-72 (2015)
[\href{https://arxiv.org/abs/1411.4413}{arXiv:1411.4413 [hep-ex]}].

\bibitem{HFLAV:2014fzu}
Y.~Amhis \textit{et al.} [HFLAV],
[\href{https://arxiv.org/abs/1412.7515}{arXiv:1412.7515 [hep-ex]}].

\bibitem{ATLAS:2012yve}
G.~Aad \textit{et al.} [ATLAS],
Phys. Lett. B \textbf{716}, 1-29 (2012)
[\href{https://arxiv.org/abs/1207.7214}{arXiv:1207.7214 [hep-ex]}].

\bibitem{CMS:2012qbp}
S.~Chatrchyan \textit{et al.} [CMS],
Phys. Lett. B \textbf{716}, 30-61 (2012)
[\href{https://arxiv.org/abs/1207.7235}{arXiv:1207.7235 [hep-ex]}].

\bibitem{Slavich:2020zjv}
P.~Slavich, S.~Heinemeyer, E.~Bagnaschi, H.~Bahl, M.~Goodsell, H.~E.~Haber, T.~Hahn, R.~Harlander, W.~Hollik and G.~Lee, \textit{et al.}
Eur. Phys. J. C \textbf{81}, no.5, 450 (2021)
[\href{https://arxiv.org/abs/2012.15629}{arXiv:2012.15629 [hep-ph]}].

\bibitem{Allanach:2004rh}
B.~C.~Allanach, A.~Djouadi, J.~L.~Kneur, W.~Porod and P.~Slavich,
JHEP \textbf{09}, 044 (2004)
[\href{https://arxiv.org/abs/hep-ph/0406166}{arXiv:hep-ph/0406166 [hep-ph]}].

\bibitem{PICO:2019vsc}
C.~Amole \textit{et al.} [PICO],
Phys. Rev. D \textbf{100}, no.2, 022001 (2019)
[\href{https://arxiv.org/abs/1902.04031}{arXiv:1902.04031 [astro-ph.CO]}].

\bibitem{IceCube:2016dgk}
M.~G.~Aartsen \textit{et al.} [IceCube],
Eur. Phys. J. C \textbf{77}, no.3, 146 (2017)
[erratum: Eur. Phys. J. C \textbf{79}, no.3, 214 (2019)]
[\href{https://arxiv.org/abs/1612.05949}{arXiv:1612.05949 [astro-ph.HE]}].

\bibitem{Planck:2018nkj}
N.~Aghanim \textit{et al.} [Planck],
Astron. Astrophys. \textbf{641}, A1 (2020)
[\href{https://arxiv.org/abs/1807.06205}{arXiv:1807.06205 [astro-ph.CO]}].

\bibitem{Djouadi:2005dz}
A.~Djouadi, M.~Drees and J.~L.~Kneur,
Phys. Lett. B \textbf{624}, 60-69 (2005)
[\href{https://arxiv.org/abs/hep-ph/0504090}{arXiv:hep-ph/0504090 [hep-ph]}].

\bibitem{ATLAS:2022hbt}
G.~Aad \textit{et al.} [ATLAS],
JHEP \textbf{06}, 031 (2023)
[\href{https://arxiv.org/abs/2209.13935}{arXiv:2209.13935 [hep-ex]}].

\bibitem{ATLAS:2024fub}
G.~Aad \textit{et al.} [ATLAS],
JHEP \textbf{05}, 150 (2024)
[\href{https://arxiv.org/abs/2402.00603}{arXiv:2402.00603 [hep-ex]}].

\bibitem{Djouadi:2002ze}
A.~Djouadi, J.~L.~Kneur and G.~Moultaka,
Comput. Phys. Commun. \textbf{176}, 426-455 (2007)
[\href{https://arxiv.org/abs/hep-ph/0211331}{arXiv:hep-ph/0211331 [hep-ph]}].

\bibitem{Kneur:2022vwt}
J.~L.~Kneur, G.~Moultaka, M.~Ughetto, D.~Zerwas and A.~Djouadi,
Comput. Phys. Commun. \textbf{291}, 108805 (2023)
[\href{https://arxiv.org/abs/2211.16956}{arXiv:2211.16956 [hep-ph]}].

\bibitem{Belanger:2018ccd}
G.~B\'elanger, F.~Boudjema, A.~Goudelis, A.~Pukhov and B.~Zaldivar,
Comput. Phys. Commun. \textbf{231}, 173-186 (2018)
[\href{https://arxiv.org/abs/1801.03509}{arXiv:1801.03509 [hep-ph]}].

\bibitem{OHare:2021utq}
C.~A.~J.~O'Hare,
Phys. Rev. Lett. \textbf{127}, no.25, 251802 (2021)
[\href{https://arxiv.org/abs/2109.03116}{arXiv:2109.03116 [hep-ph]}].

\bibitem{ATLAS:2023xco}
 [ATLAS],
[\href{https://atlas.web.cern.ch/Atlas/GROUPS/PHYSICS/PUBNOTES/ATL-PHYS-PUB-2023-025/fig_15.png}{ATL-PHYS-PUB-2023-025: FIG. 15}].

\bibitem{ATLAS:2014zve}
G.~Aad \textit{et al.} [ATLAS],
JHEP \textbf{05}, 071 (2014)
[\href{https://arxiv.org/abs/1403.5294}{arXiv:1403.5294 [hep-ex]}].

\bibitem{ATLAS:2019lng}
G.~Aad \textit{et al.} [ATLAS],
Phys. Rev. D \textbf{101}, no.5, 052005 (2020)
[\href{https://arxiv.org/abs/1911.12606}{arXiv:1911.12606 [hep-ex]}].

\bibitem{ATLAS:2023djh}
 [ATLAS],
[\href{https://cds.cern.ch/record/2861058?ln=en}{ATLAS-CONF-2023-029}].

\bibitem{CMS:2020bfa}
A.~M.~Sirunyan \textit{et al.} [CMS],
JHEP \textbf{04}, 123 (2021)
[\href{https://arxiv.org/abs/2012.08600}{arXiv:2012.08600 [hep-ex]}].

\bibitem{CMS:2023qhl}
 [CMS],
[\href{https://cds.cern.ch/record/2853345?ln=en}{CMS-PAS-SUS-21-008}].

\bibitem{CMS:2022syk}
A.~Tumasyan \textit{et al.} [CMS],
Phys. Rev. D \textbf{108}, no.1, 012011 (2023)
[\href{https://arxiv.org/abs/2207.02254}{arXiv:2207.02254 [hep-ex]}].

\end{thebibliography}
\end{document}